\documentclass[useAMS,usenatbib,usegraphicx]{mn2e}
\usepackage{epsfig,amssymb}
\usepackage[dvips]{color}

\newcommand{\Msun}{M_{\odot}}

\definecolor{Black}{named}{Black}
\definecolor{Red}{named}{Red}

\title[O VII column densities from simulations]{Hot gas haloes around disc galaxies: 
O VII column densities from galaxy formation simulations}

\author[E. Ntormousi and J. Sommer-Larsen]{E. Ntormousi$^{1,2}$\thanks{E-mail:
eva@usm.uni-muenchen.de (EN); jslarsen@astro.ku.dk (JSL)} and J. Sommer-Larsen$^{3,4,5}$\\
$^{1}$Universit\"{a}ts-Sternwarte, Munich D-81679, Germany\\
$^{2}$Max-Planck Institut f\"{u}r Extraterrestrische Physik, 85741 Garching
Germany\\
$^{3}$Dark Cosmology Center, Niels Bohr Institute, University of Copenhagen, 
Juliane Maries Vej 30, DK-2100 Copenhagen, Denmark\\
$^{4}$Excellence Cluster Universe, Boltzmannstr. 2
85748 Garching, Germany\\
$^{5}$Marie Kruses Skole, Stavnsholtvej 29-31, DK-3520 Farum, Denmark}

\begin{document}

   \date{Accepted 2010 July 13.  Received 2010 July 2; in original form 2010 April 9}

   \pagerange{\pageref{firstpage}--\pageref{lastpage}} \pubyear{2009}

   \maketitle

   \label{firstpage}

   \begin{abstract}

Numerical models of disc galaxy formation predict the existence of extended, hot ($T\sim10^6$K) gas haloes around present day spirals.  The X-ray luminosity of these haloes is predicted to increase strongly with galaxy mass.  However, searches for their X-ray emission have not been successful so far.

We calculate the all sky O VII column density distributions for the haloes of three Milky Way like disc galaxies, resulting from cosmological high-resolution, N-body/gasdynamical simulations.  We perform calculations both including the disc gas and without it, so the disc contribution to the column density is quantified.  It is found that the column densities estimated for Milky Way-like galaxies are just below the observational upper limit, making a test of the hot halo paradigm likely within observational reach.
 
\end{abstract}


   \section{Introduction}

%%%%%%%%%%%%%%%%%%%%%%%%% Table 1 %%%%%%%%%%%%%%%%%
\begin{table*}
\caption{Numerical and physical  
characteristics of the galaxy simulations at $z$=0: mass of DM/(high-res)gas/star particles and the respective gravitational (spline) softening
lengths; total number of particles,  total number of SPH particles and total number of SPH 
particles after removal of disc etc.; virial mass and radius of the galaxy halo; characteristic circular speed of the galaxy}
\begin{tabular}{l c c c c c c c c c c c c c}
\hline
run &   $m_{DM}$ &  $m_{gas}$ & $m_{*}$ & $\epsilon_{DM}$  &  $\epsilon_{gas}$  &  $\epsilon_{*}$  &
$N_{tot}$ & $N_{SPH}$ & $N_{SPH}'$ & $M_{vir}$  &  $R_{vir}$ & $V_c$ \\
        &  & [$10^6 M_{\odot}/h$] &  &  & [kpc/$h$] &  &  & & & [$10^{12}~M_{\odot}$] &
        [kpc] & [km/s] \\
\hline  
K33      & 0.52 & 0.093 & 0.093 & 0.34 & 0.19 & 0.19 & 1120000 & 291000 & 227000 & 0.39 & 196 & 180 \\
K26   & 4.2 & 0.012 & 0.73 & 0.68 & 0.095 & 0.38 & 733000 & 534000 & 333000 & 0.61 & 230 & 207 \\
K15      & 4.2 & 0.012 & 0.73 & 0.68 & 0.095 & 0.38 & 780000 & 552000 & 458000 & 0.90 & 262 & 245\\
K33LR   & 4.2 & 0.012 & 0.73 & 0.68 & 0.095 & 0.38 & 537000 & 425000 & 257000 & 0.39 & 196 & 180 \\
\hline
\end{tabular}
\label{tab:data}
\end{table*}
%%%%%%%%%%%%%%%%%%%%%%%%%%%%%%%%%%%%%%%%%%%%%%%%%%%%%
        
Numerical simulations of disc galaxy formation predict that disc galaxies, not only in groups or clusters, but also in isolation, should reside in extended, hot gaseous haloes ($T\sim10^6$ K).   This gas is accreted from the intergalactic medium and cools out as it settles onto the disc, thus contributing to the formation of the galaxy.  The mass of the predicted halo gas is comparable to the total baryonic mass of the galactic disc and could be the answer to the 'missing galactic baryons' problem \citep{SL06}.

Gas at temperatures $T\sim10^6$ K is best studied in the X-ray regime.  The X-ray luminosities of the galactic haloes are predicted to depend strongly on the characteristic circular velocity of the galaxy, $L_X\propto V_c^7$ \citep{toft2002} and should be detectable for the most massive galaxies.  However, searches for this X-ray emitting gas have so far been unsuccessful for quiescent disc galaxies. Given the $L_X\propto V_c^7$ scaling, the most restrictive test
has been been performed by \citet{rasmussen09}, who observed the halo of the 
$V_C = 307$ km/s, quiescent, edge-on disc galaxy NGC 5746, and obtained an 
upper limit compatible with, but not far from, theoretical predictions.

Plenty of evidence of a hot dilute halo around the Milky Way exists at
present: 1) the head-tail structure of many of the  high-velocity HI 
clouds (HVCs) indicates that they are moving in a medium much more hot and 
dilute than themselves \citep{bruns2000, ganguly05}, 2) the absence of HI in Local
Group dwarf galaxies within $\sim$270 kpc distance of either the Milky Way
or M31 \citep{GP09}, 3) the Magellanic Stream \citep{stanim2008}, given the revised orbit
of the LMC \citep{M09} and 4) high-velocity O VI line absorption in quasar 
spectra \citep{sembach03, wakker05}. O VI probes gas at temperatures near $10^5$ K, 
so high-velocity O VI associated with hydrogen indicates a thermally 
unstable interface between the high velocity cool clouds and a hot 
surrounding medium. Moreover, cool-out of hot halo gas, at a rate of
$\sim$~1 $\Msun$/yr is indicated by SiIII absorption by high- and
intermediate-velocity clouds in the spectra of AGNs \citep{Sh.09}.
Finally, the warm-hot phase of the intergalactic medium \citep{nicastro05} provides a reservoir of hot and dilute gas at galactic distances.
       
       Six times ionized oxygen is indicative of temperatures around $10^6$ K so, considering the high cosmic oxygen abundance, the O VII line is the best way to check for the existence of a hot gas halo around the Milky Way on the basis
of absorption line studies. Absorption in X-rays has been routinely observed at zero redshift for some time \citep{nicastro02, rasmussen03, fang06}, but the question of where exactly this gas is located remains under study.  \citet{bregmanlloyd07} compared X-ray absorption between lines of sight in different directions on the sky, and concluded a galactic halo origin for the absorbing gas is plausible.  \citet{yao2008} made a search for the X-ray halo of the Milky Way by comparing O VII line absorption between spectra of objects at different distances and
positions relative to the galactic plane. The difference between absorption in sightlines that mostly probe the disc and others that also probe the
halo provides an upper limit on
the contribution of the halo gas to the total absorption. From this study,
an upper limit of $N(O VII)<5\cdot10^{15}$ cm$^{-2}$ was found for the
sightline considered through the galactic halo.  \citet{yao2010} calculated a somewhat lower column-density contribution of the putative galactic haloes to X-ray abroption by stacking observations towards AGNs and using the intervening galaxies as tracers. Their study gave an upper limit to the ionic column densities coming from galaxy haloes ranging from $log(N(O VII))<14.2-14.8$, but we note that
these constraints are less restrictive, since the sight-lines will, in general,
pass through the haloes at considerable distances from the galaxy centres - see
also the discussion in section 4.

       In this paper we show that there is no discrepancy between these observational limits and predictions from galaxy formation simulations. Given information on gas density, temperature, oxygen abundance
and UVB radiation field from the simulations, we use the (photo-)ionization
code CLOUDY \citep{ferland98} to determine the O~VII column density along
all lines-of-sight (from positions in the disc corresponding to the location
of the sun in the Galactic disc) in the simulated galaxies.
We perform the analysis with and without the contribution from the disc gas
included.

       The structure of this paper is as follows:  In section \ref{num_method} we describe the method we used for our calculations, in section \ref{results} we present the results, and in section \ref{discussion} we discuss the results
obtained.

    \section{Numerical Method}\label{num_method}

       \subsection{The N-body/gasdynamical galaxy formation simulations}

          For our analysis we used three simulated, Milky Way like disc galaxies, 
resulting from cosmological galaxy formation and evolution simulations based 
on a flat $\Lambda CDM$ cosmology ($\Omega_m=0.3, \Omega_{\Lambda}=0.7$).
 
The code used for the simulations was a significantly improved version of
the TreeSPH code, which has been used previously for galaxy formation 
simulations (\cite{SLGP03}, in the following SLGP).
The main improvements over the previous version are:
(1) The ``conservative'' entropy 
equation solving scheme suggested by \cite{SH02} has been adopted. 
(2) Non-instantaneous gas recycling and chemical evolution, tracing
10 elements (H, He, C, N, O, Mg, Si, S, Ca and Fe), has been incorporated
in the code following Lia et~al.\ (2002a,b); the algorithm includes 
supernov\ae\ of type II and type Ia, and mass loss from stars of all masses.
(3) Atomic radiative cooling depending both on the metal abundance
of the gas and on the meta--galactic UV field, modelled after \cite{HM96}
is invoked, as well as simplified treatment
of radiative transfer, switching off the UV field where the gas
becomes optically thick to Lyman limit photons on scales of $\sim$ 1~kpc.

Simulations of three disc galaxies, of mass and size comparable to that
of the Milky Way, were used for the present work.
The galaxies selected, denoted K33, K26 and K15 in the following, 
represent ``field'' galaxies (SLGP), and have, 
at $z$=0, characteristic circular speeds of $V_c= 180, 207$ and 245 km/s, and 
virial masses of 0.4-0.9x10$^{12} \Msun$.

The galaxies (galaxy DM haloes) were initially selected from a
cosmological, DM-only simulation of box-length 10 $h^{-1}$ Mpc
(comoving), and starting redshift $z_i$=39.  
Mass and force resolution was increased in Lagrangian regions enclosing the 
galaxies, and in these regions all DM particles were split into a DM particle
and a gas (SPH) particle according to an adopted universal baryon fraction of
$f_b$=0.15, in line with recent estimates. Particle numbers for the
three TreeSPH simulations were in the range 0.3-1.1x10$^6$ (comparison
to a lower resolution simulation consisting of 1.5x10$^5$ particles
will be briefly made).

For the galaxy of $V_c = 180$ km/s, $m_{\rm{gas}}$=$m_*$=
9.3x10$^4$ and $m_{\rm{DM}}$=5.2x10$^5$ $h^{-1}$M$_{\odot}$.
Moreover, gravitational (spline) 
softening lengths of $\epsilon_{\rm{gas}}$=$\epsilon_*$=190 and 
$\epsilon_{\rm{DM}}$=340 $h^{-1}$pc, respectively, were adopted. 
For the galaxies of $V_c$~=~207 and 245 km/s, $m_{\rm{gas}}$=$m_*$=
7.3x10$^5$ and $m_{\rm{DM}}$=4.2x10$^6$ $h^{-1}$M$_{\odot}$, and
$\epsilon_{\rm{gas}}$=$\epsilon_*$=380 and 
$\epsilon_{\rm{DM}}$=680 $h^{-1}$pc. The
gravity (spline) softening lengths were fixed in physical coordinates 
from $z$=6 to $z$=0, and in comoving coordinates at earlier times.

A Kroupa IMF was used in the simulations, and early rapid and 
self-propagating star-formation (sometimes dubbed ``positive feedback'')
was invoked (SLGP). Finally, in order to enable some reuse of previous
work, values of $h$=0.65 and $\sigma_8$=1.0 were employed in the 
cosmological simulations. Test simulations adopting $h=0.7$ and 
$\sigma_8=0.9$ give very similar results, since they produce in total 
only about $5-10\%$ more hot gas \citep{SL06}.

In order to increase the resolution of the (dilute) halo gas, and enable
the formation of a halo multi-phase medium (though still dominated by
the hot gas phase; Sommer-Larsen 2006), for galaxies K15 and K26
(which were run at eight times lower mass resolution and two times lower
force resolution than K33), each gas particle of $T>20000$ K, was split into
8 particles 0.7 Gyr before the end of the simulation 
(i.e., $z$=0). The simulations were then continued for 0.2 Gyr, and,
for galaxies K26 and K33, each gas particle of $T>20000$ K
was then again 
split into 8 gas particles. The resulting ultra-high resolution gas particle 
mass was 1.2x10$^4$~$h^{-1}\Msun$, and gravity softening length 
$\epsilon_{\rm{gas}}$=95$h^{-1}$pc for the two simulations, which were
then continued for an additional 0.5 Gyr, till $z$=0. This allows the
development of a multi-phase halo gas structure, including neutral
gas clouds (``high-velocity clouds'') --- more detail is given in
\cite{SL06} and \cite{peek2008}.

Finally, to test for effects of numerical resolution, a simulation
of galaxy K33, but at the (lower) resolution of galaxies K26 and
K15, was also run. Moreover, as detailed above, gas particles of $T>20000$ K were split into
8 particles 0.7 Gyr before the end of the simulation and yet
again split into 8 particles 0.5 Gyr before $z$=0.

      \subsection{Oxygen fractions from CLOUDY}\label{cloudy}

          We calculated the fraction of oxygen in each ionisation state using CLOUDY version 7.2, last described by \shortcite{ferland98}.  CLOUDY is a spectral synthesis code, designed to simulate astronomical plasmas. 

          We used the code for a plane-parallel geometry setup, assuming constant temperature of the gas and including a background UV ionizing field from \shortcite{hm1997}, adapted to $z=0$ as described in CLOUDY's user guide.  The calculation was done for a single zone and iteration was allowed to continue until convergence.  We thus created a look-up table of O VII fractions for hydrogen densities ranging from $\log{n_H}=-10$ to $\log{n_H}=-3$ and gas temperatures from $\log{T}=3.5$ to $\log{T}=7$.  The gas was assumed to be optically thin to the UV radiation.  Self shielding of the cold and dense gas was approximated by "turning off" the UV background for temperatures $T<20000$ K and densities $n_{H}>10^{-3}$ cm$^{-3}$ --- this condition, at $z\sim0$, effectively mimics the (simplified) UVB radiative transfer invoked in the galaxy formation simulations.

\begin{figure}
            %\vspace*{174pt}
            \centering
            \caption{O~VII fractions, $f_7$, calculated using CLOYDY and invoking 
a $z$=0 Haardt \& Madau background UV field. The effect of the UV field 
becomes significant at hydrogen densities $n_H\la10^{-4}$ cm$^{-3}$}
            \epsfig{file = 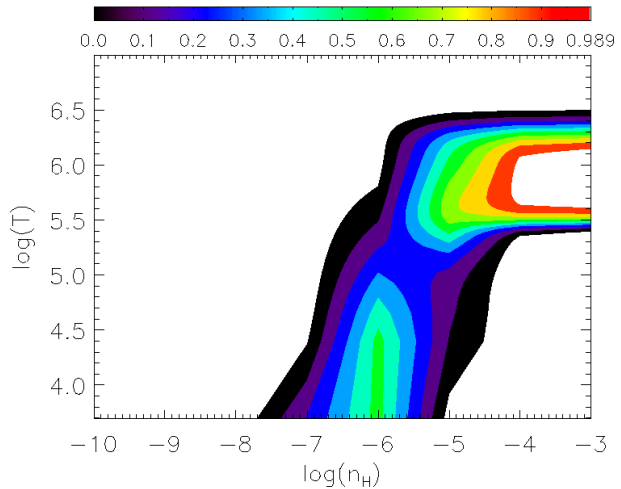, width = 0.5\textwidth}
            \label{fractions}
          \end{figure}

          Without a background radiation field and for thermodynamical equilibrium, ionisation fractions only depend on the gas temperature, according to the Saha equation \citep{saha1921}.  However, with the inclusion of the UV field, the 
ionization state also depends on the density. 
The dependence of the O VII fraction on density and temperature,
at $z$=0, is shown in Figure \ref{fractions}.

      \subsection{Calculating O VII column densities}\label{calc}

         For our analysis we used SPH simulation data for three spiral galaxies.  We calculated column densities as,

          \begin{equation}\label{colden}
             N(O VII) = \sum^{N}{n_{OVII}\cdot d} ~,
          \end{equation}  
with
          \begin{equation}\label{colden1}
             n_{OVII} = n_H\cdot Z_{O,sim}\cdot Z_{O,\odot}\cdot f_7 ~,
          \end{equation}  
where $n_H$ is the hydrogen number density in cm$^{-3}$, $Z_{O,sim}$ is the oxygen abundance in units of solar, $Z_{O,\odot}$ is the solar oxygen number abundance from \shortcite{asplund05}, d the distance in cm between two points on a line of sight, $f_7$ is the fraction of oxygen in the sixth ionisation state and the sum is over a set of equidistant points along a given line of sight.  $n_H$ and $Z_{O,sim}$ were given by the simulations, while the oxygen fractions were calculated using CLOUDY, as described in section \ref{cloudy}.

For each of the simulated galaxies we drew lines of sight from within the disc into the halo, covering all the sky.  
The starting point of each line was placed on the midplane of the disc, approximately 8 kpc from the centre and the integration was carried into the halo, up to a distance of 500kpc.  The choice of 500kpc is discussed in section \ref{results}.

Each line of sight was represented by 150 equidistant points and to each point $\vec{p}$, a smoothing length was assigned,         
         \begin{equation}
           \label{sp}
           h_{\vec{p}}=\frac{1}{4}(d_{50}+d_{51}) ~~,
         \end{equation}
         where $h_{\vec{p}}$ is the smoothing length of the line point and $d_{50}$ and $d_{51}$ are the distances of the 50th and 51st closest SPH particles to the point, respectively.  This choice was made for consistence with the SPH calculations, which use the 50 closest neighbours of a particle to evaluate interaction forces.
All quantities of interest were then averaged over a sphere of radius 2$h_{\vec{p}}$,
centred on the point $\vec{p}$, using a cubic spline weighting function that mimics the SPH kernel function, and attributed to the point.  

Specifically, for each SPH particle, on the basis of the density and 
temperature, the OVII fraction, $f_7$, was calculated using the CLOUDY based
lookup table we have created. Using the oxygen abundance, $n_{OVII}$ was
then determined. These values were then interpolated onto the individual 
line-of-sight points, as described above. Finally, eq.\ref{colden} was used
to calculate N(OVII) along a given line-of-sight.

For choosing the number of points to represent a line of sight, we compared the results of 20000 line of sight calculations, each done using 150 points and then repeated using 500 points and found only insignificant differences in the column densities.  Higher line-of-sight resolution than this is not justified, as we would then interpolate particle properties over distances smaller than the SPH smoothing lengths.    

   \section{Results}\label{results}

Not only hot halo gas may contribute to the O~VII column density along a
given line-of-sight. Also gas in the disc, heated by feedback from super-nova
explosions or photo-ionized, may contain O~VII. Furthermore, satellite
galaxies with ongoing star formation and related super-nova feedback,
as well as the photo-ionized surface layers of HI clouds in the halo (HVCs),
may contribute to the O~VII column density along some lines-of-sight.

In our first set of N(OVII) estimates, in order to exclude such 
``contaminating'' gas,
we excluded all disc gas within a cylinder of radius $r=R_d = 20$ kpc, 
where $R_d$ 
the galactic disc radius and height $z=\pm3$ kpc, centred at the centre of 
the galaxy, and aligned with the disc.
Moreover, it was found that potentially ``contaminating''
gas in satellites as well as in halo HI clouds was characterized by
densities $n_H > n_{H,c}$, where $n_{H,c}=10^{-3}$ cm$^{-3}$. On the other
hand, the hot halo gas was found to everywhere have $n_H < n_{H,c}$. We
hence excluded all gas in the halo of $n_H > n_{H,c}$ from the analysis.
The filling factor of this gas is very small, $f \ll 0.01$, and the
effect of excluding it on the results presented in the following
is negligible.

In order to assess the effect of removing the potentially contaminating
gas we made a second set of N(OVII) estimates based on all gas, i.e.,
including gas in the disc, satellite galaxies and HI clouds.

We should note here that, when choosing our lines of sight, as a first approach we did not try to achieve the same resolution 
in all areas of the celestial sphere.  
However, any over- or under- sampling of areas could result in differences in the calculated distributions.  
For this reason, we have compared between a constant 1 degree separation of the lines of sight and a simple 
weighting scheme that samples each galactic longitude with $360\cdot\cos{b}$ lines, 
where b is the galactic latitude of the line of sight, and found little difference in the statistics.  
In the following we use the constant resolution for illustration purposes only and present 
all the analysis for the data calculated with the weighting scheme.

\begin{figure*}
             \begin{minipage}{\textwidth}
                \begin{center}
                   \caption{Histograms of the number of lines of sight at each column density bin.  
                      Left column: Blue dashed histograms result from including all gas particles, while red solid histograms result from excluding the disc and ''dense'' halo objects (see text for details).
                      Right column: Red solid histograms are the same as shown in the right 
column (including a UV background and excluding the disc and ``dense'' halo
objetcs), blue dashed histograms result from excluding the UV background (see
text for details). 
                      Numbers are normalized to the total number of lines for each model. From top to bottom: K33, K26, K15}
                      \begin{tabular}{p{0.5\linewidth}p{0.5\linewidth}}
                         \epsfig{file=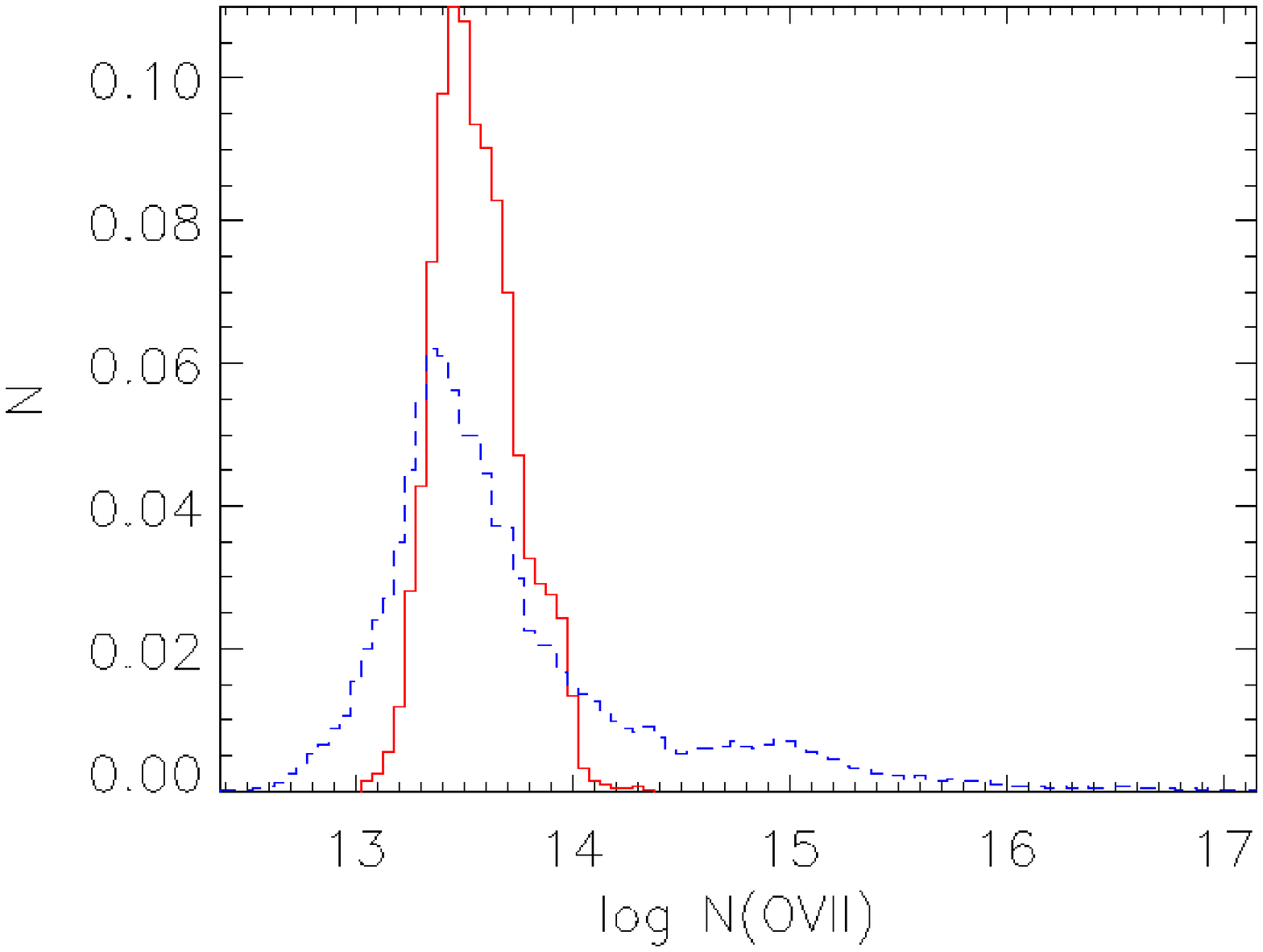,width=\linewidth} &
                         \epsfig{file=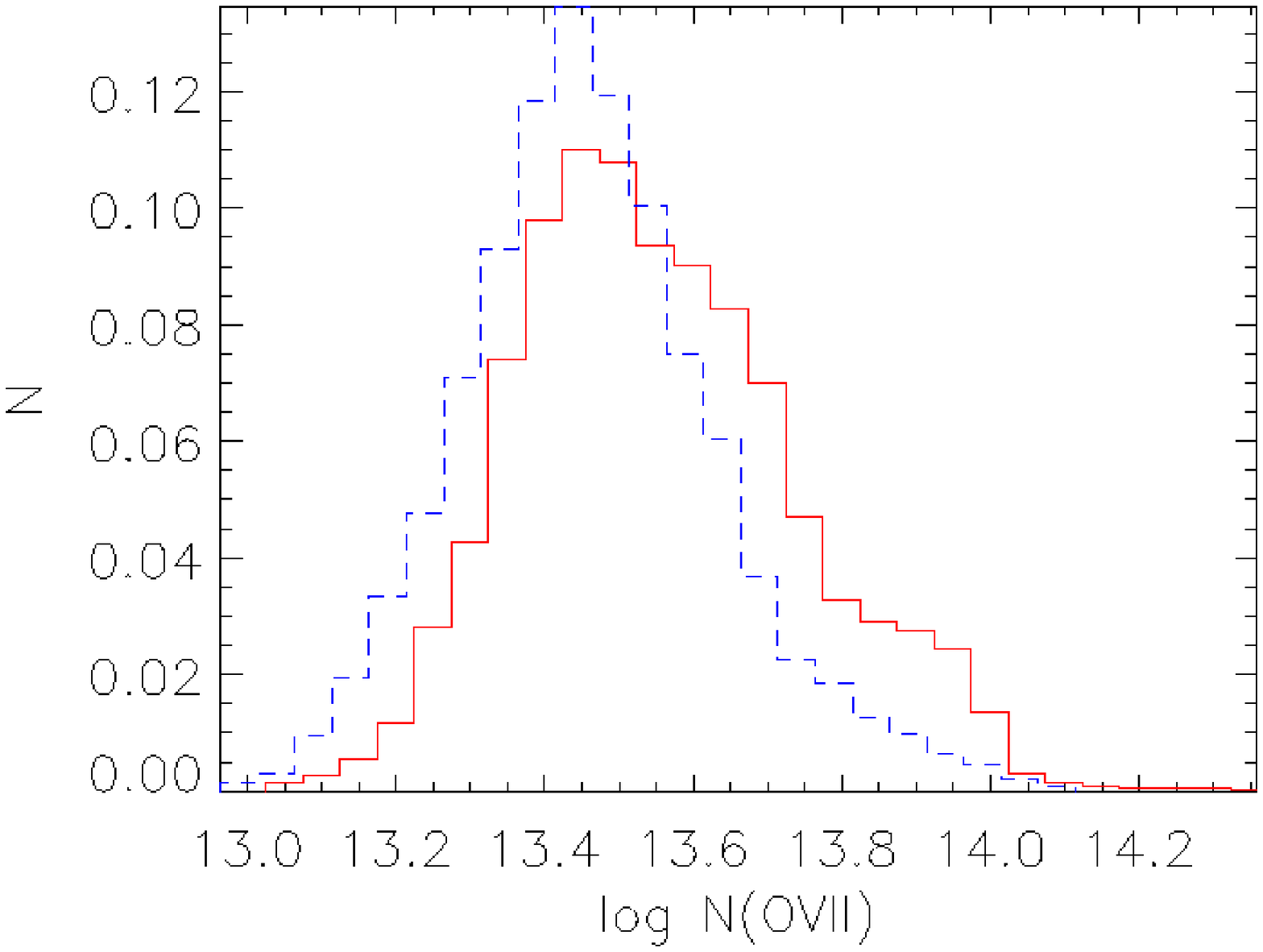,width=\linewidth} \\
                         \epsfig{file=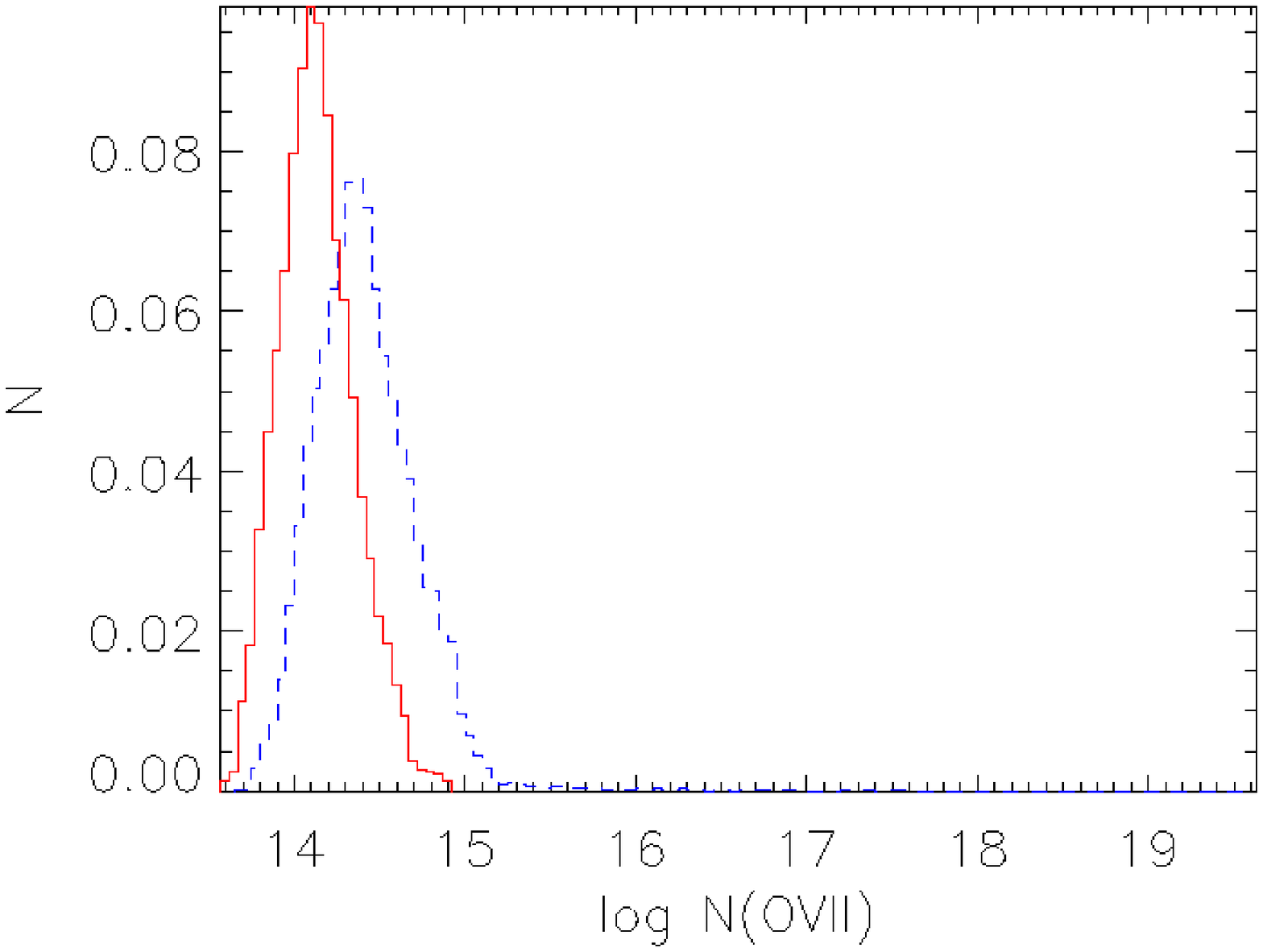,width=\linewidth} & 
                         \epsfig{file=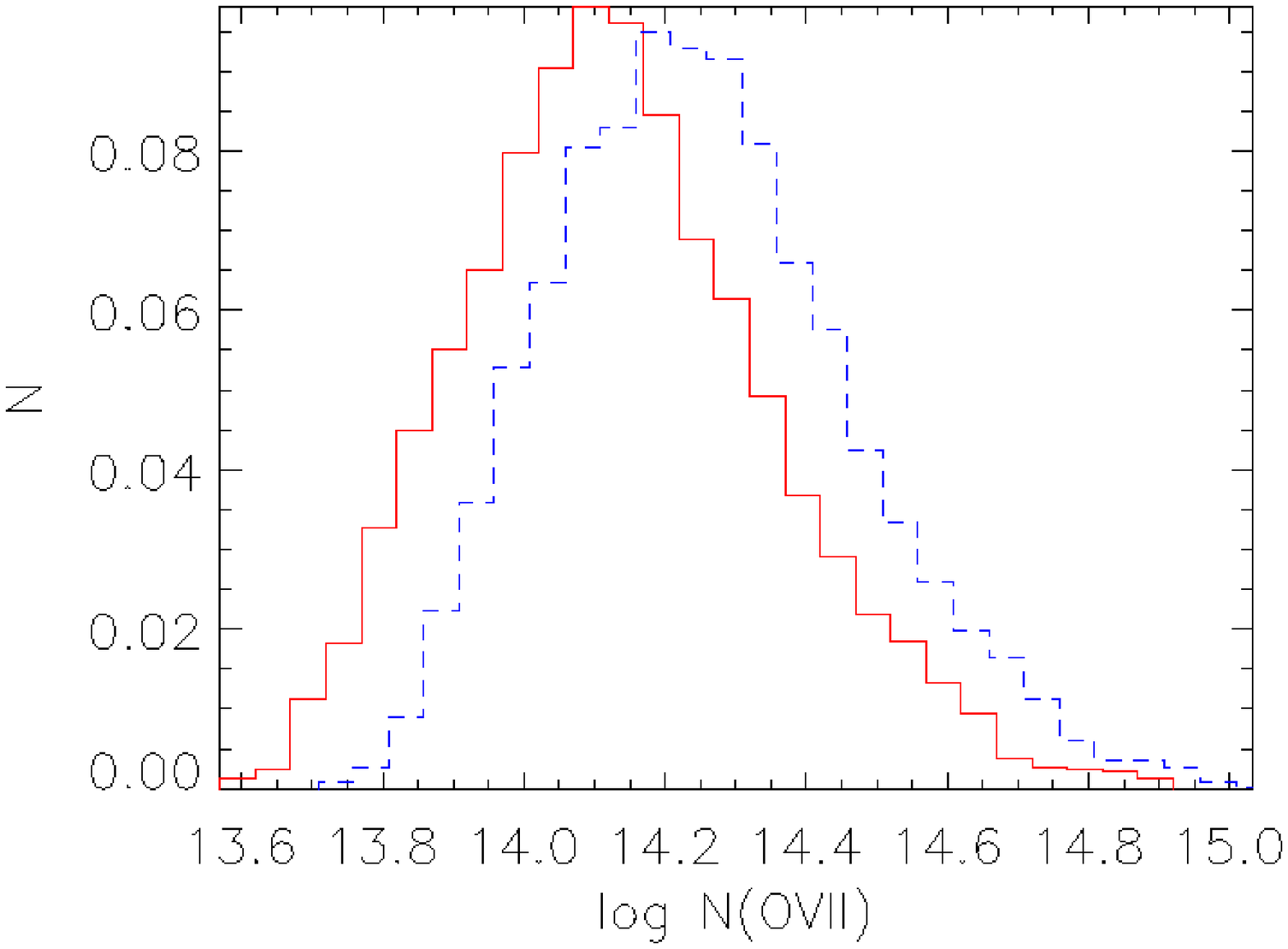,width=\linewidth} \\
                         \epsfig{file=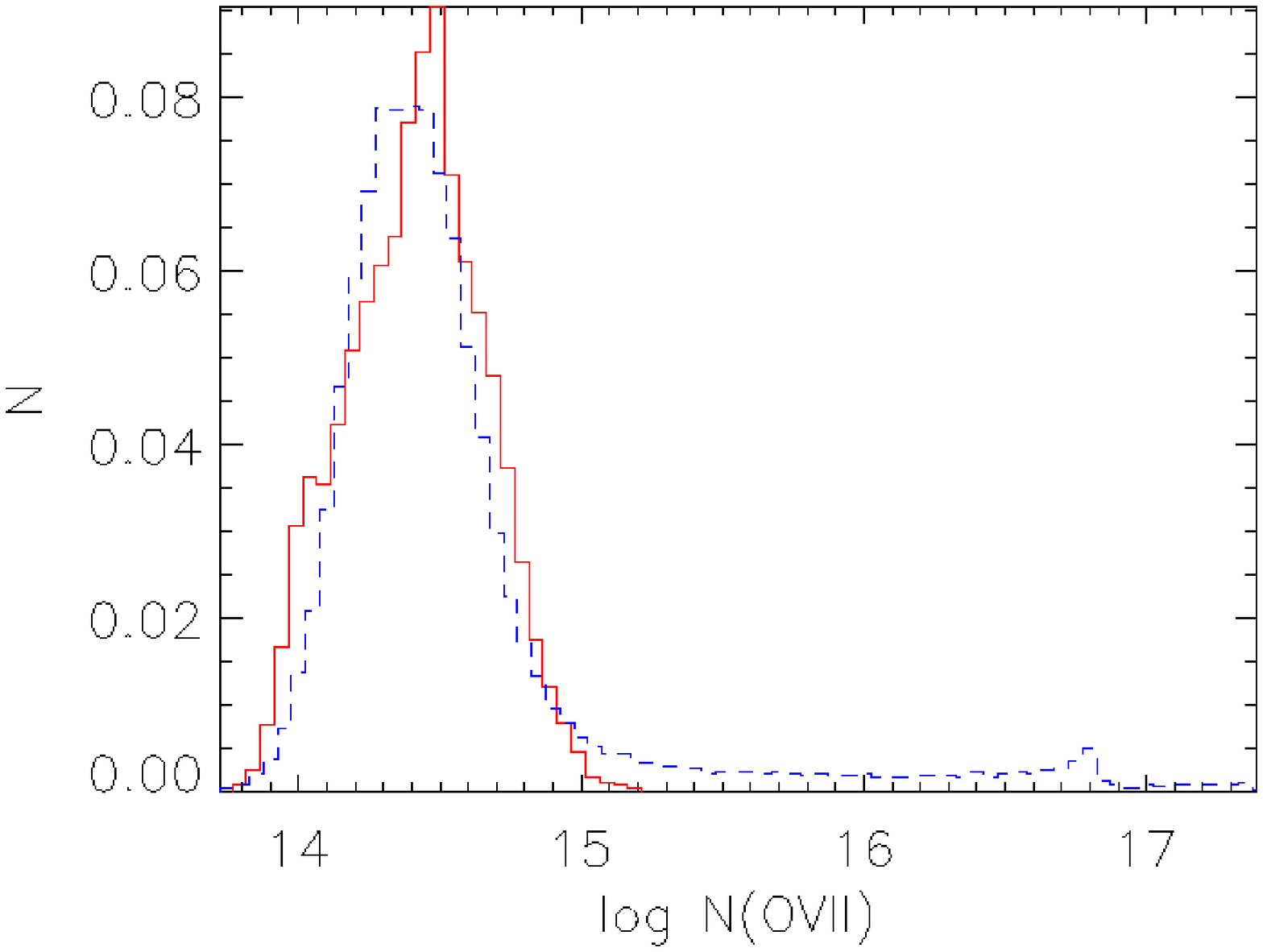,width=\linewidth} &
                         \epsfig{file=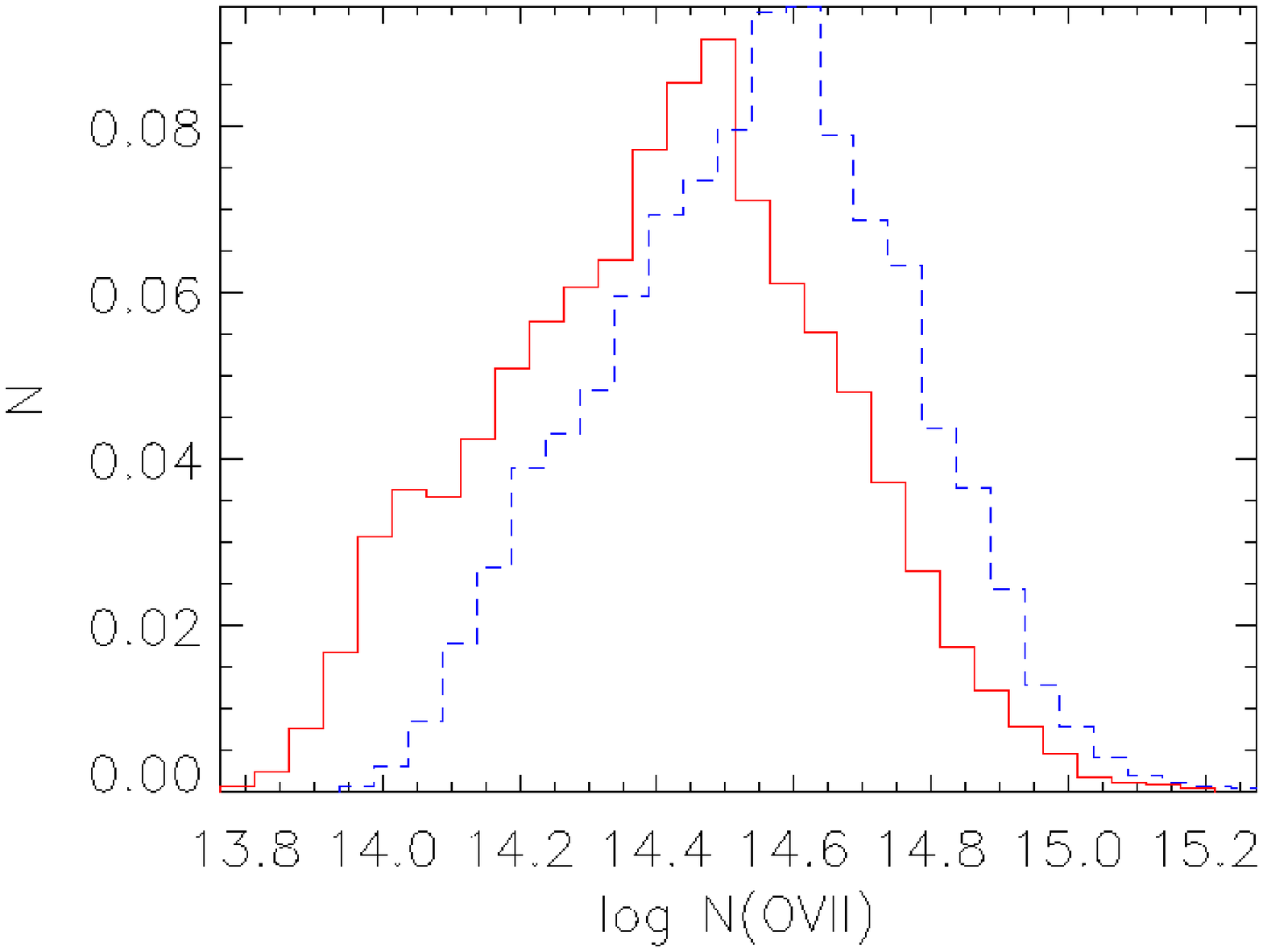,width=\linewidth}
                         \label{hist}
                      \end{tabular}
                 \end{center}
             \end{minipage}
        \end{figure*}

\begin{figure*}
             \begin{minipage}{\textwidth}
                \begin{center}
                   \caption{Cumulative histograms of the O~VII column densities (fraction of lines of sight with column densities smaller than the corresponding column density bin).  
                  Left column: Blue dashed histograms result from including all gas particles, while red solid histograms result from excluding the disc and ''dense'' halo objects (see text for details).
                      Right column: Red solid histograms are the same as shown in the right 
column (including a UV background and excluding the disc and ``dense'' halo
objetcs), blue dashed histograms result from excluding the UV background (see
text for details).    
                      From top to bottom: K33, K26, K15}
                      \begin{tabular}{p{0.5\linewidth}p{0.5\linewidth}}
                         \epsfig{file=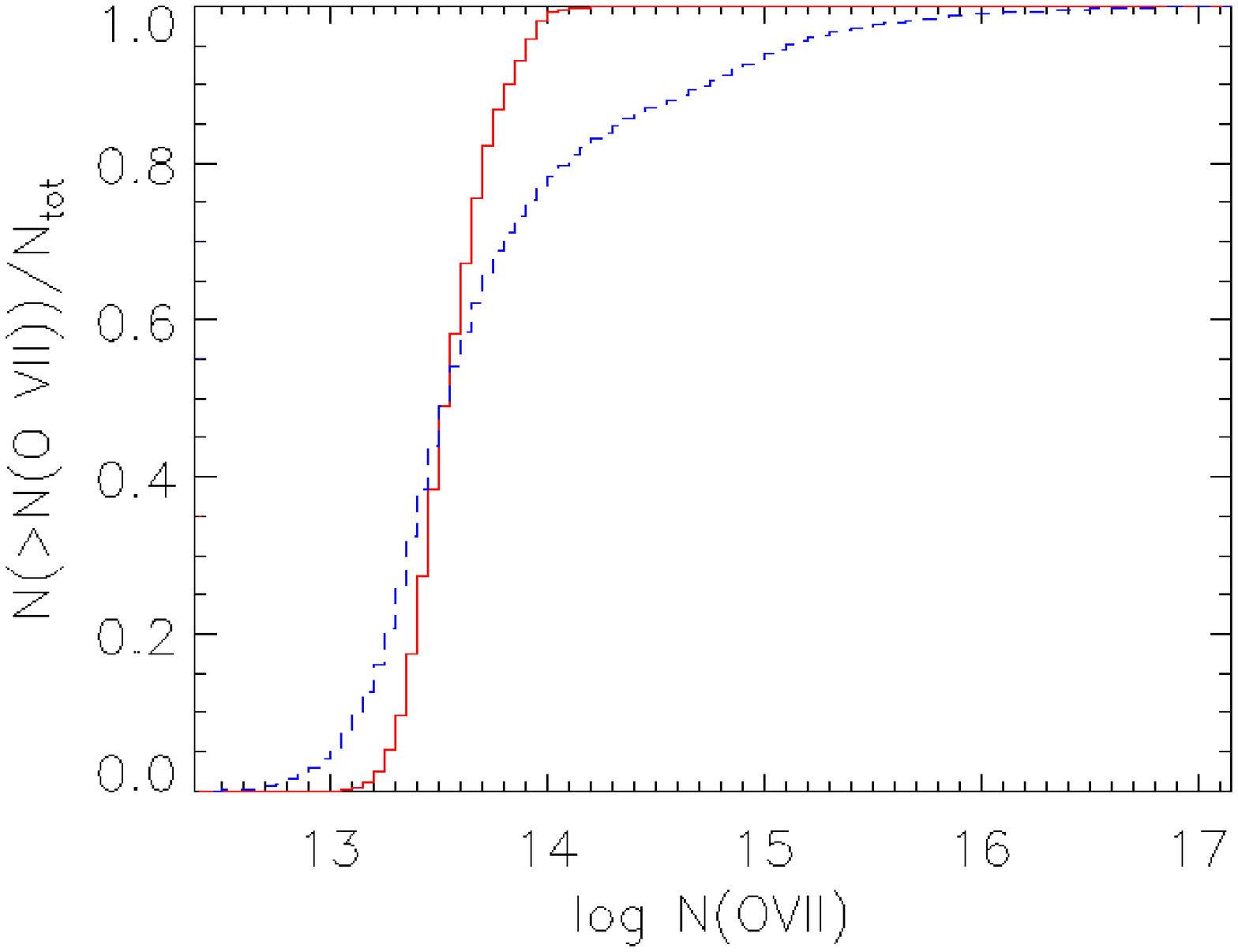,width=\linewidth} &
                         \epsfig{file=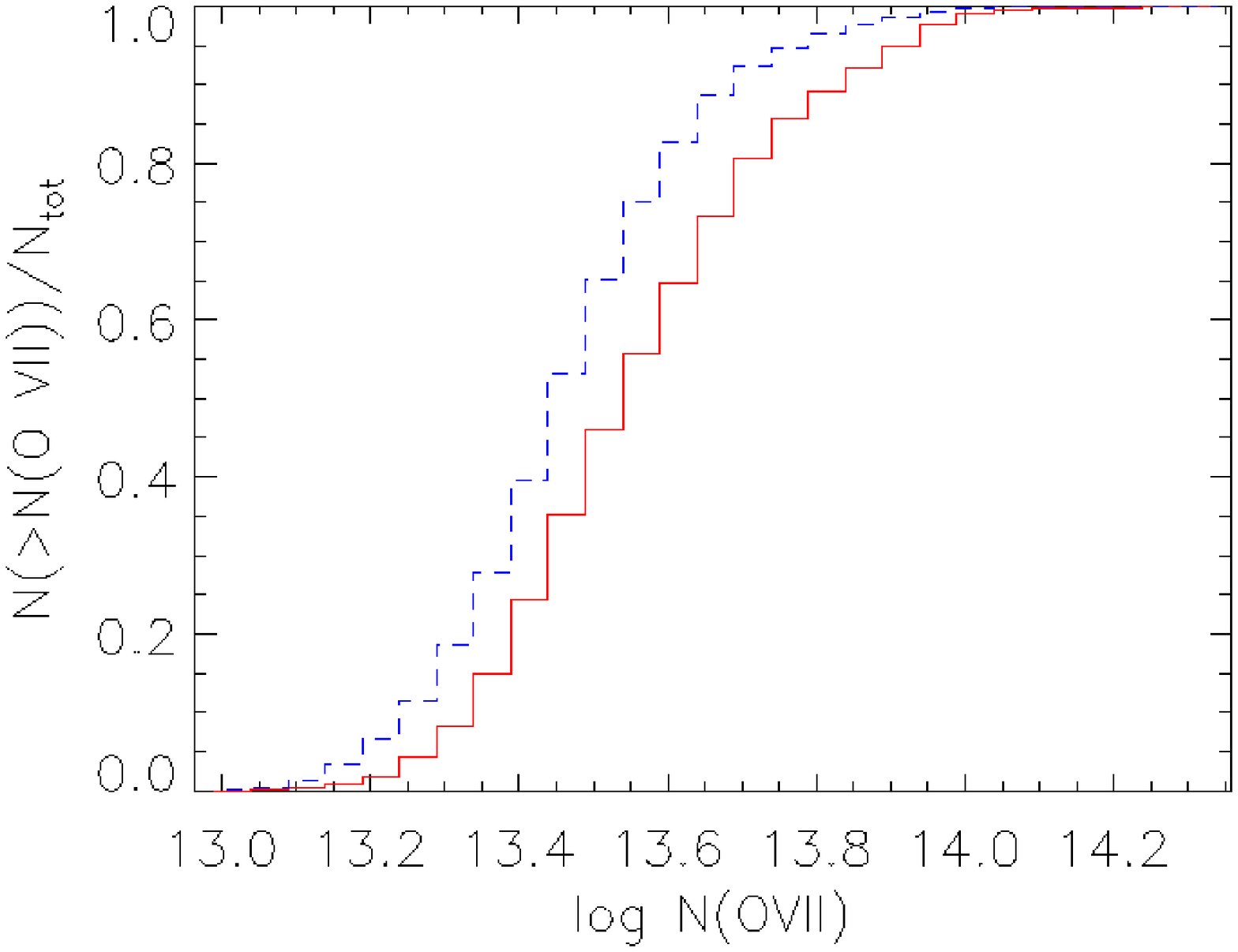,width=\linewidth} \\
                         \epsfig{file=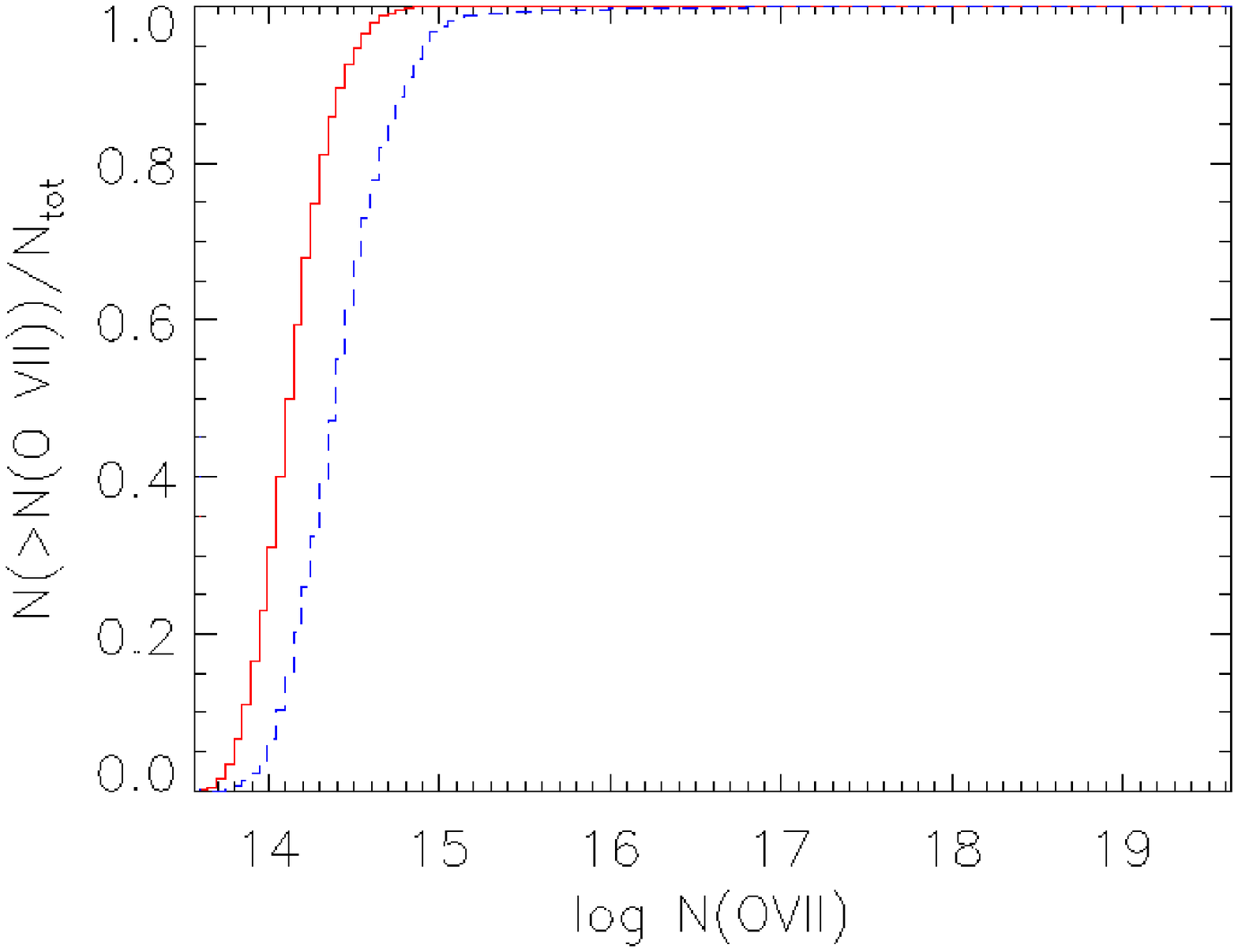,width=\linewidth} & 
                         \epsfig{file=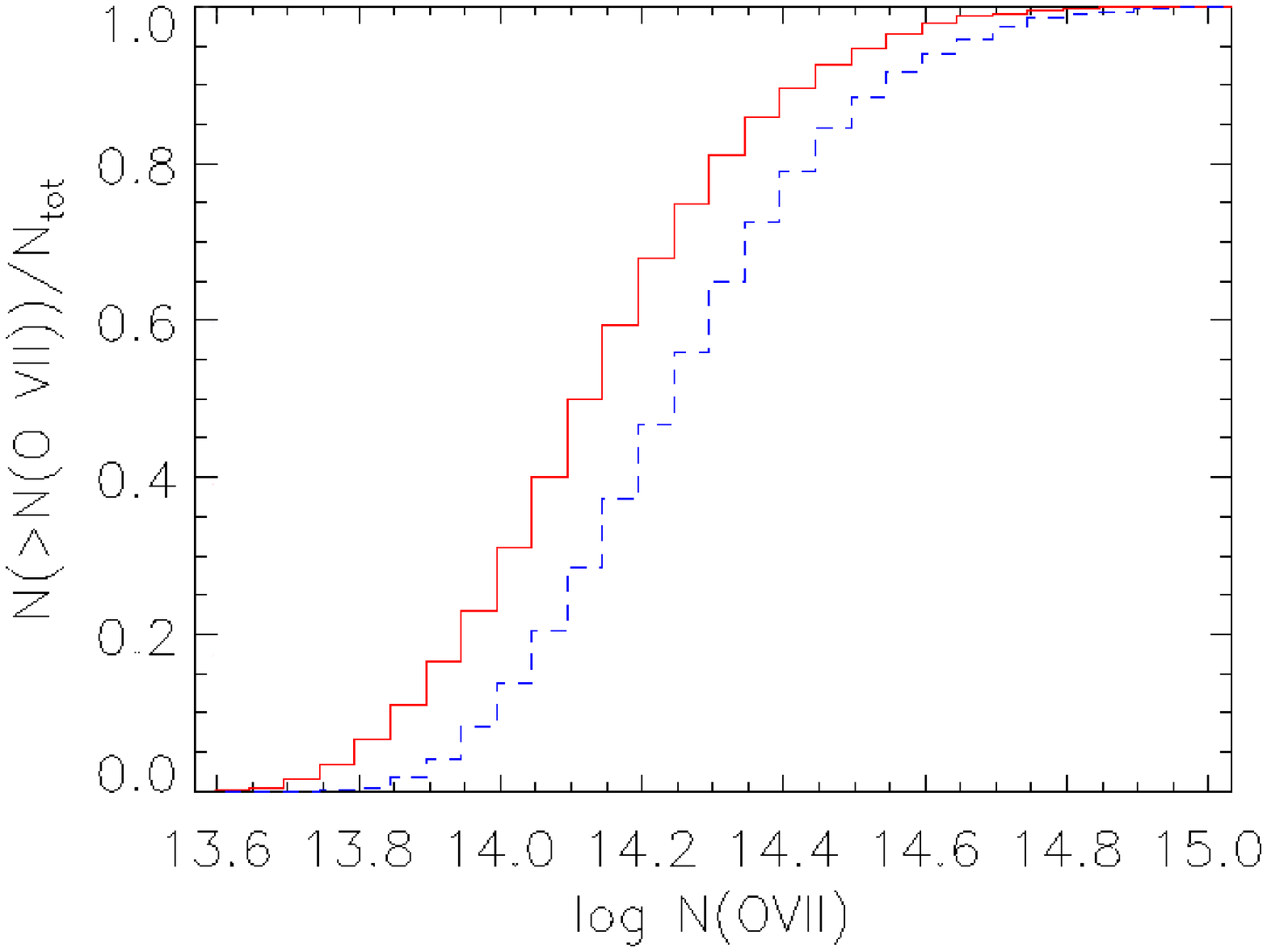,width=\linewidth} \\
                         \epsfig{file=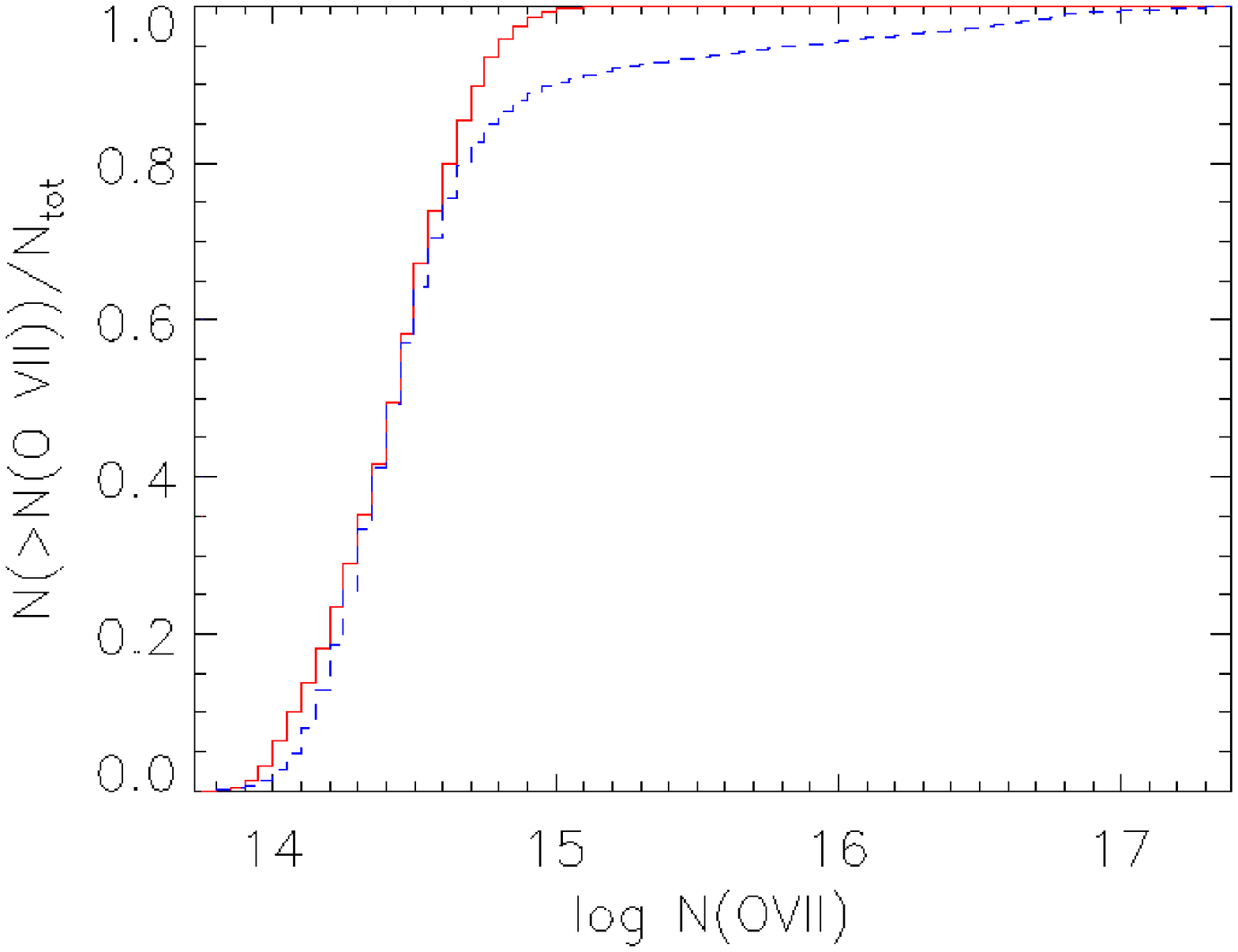,width=\linewidth} &
                         \epsfig{file=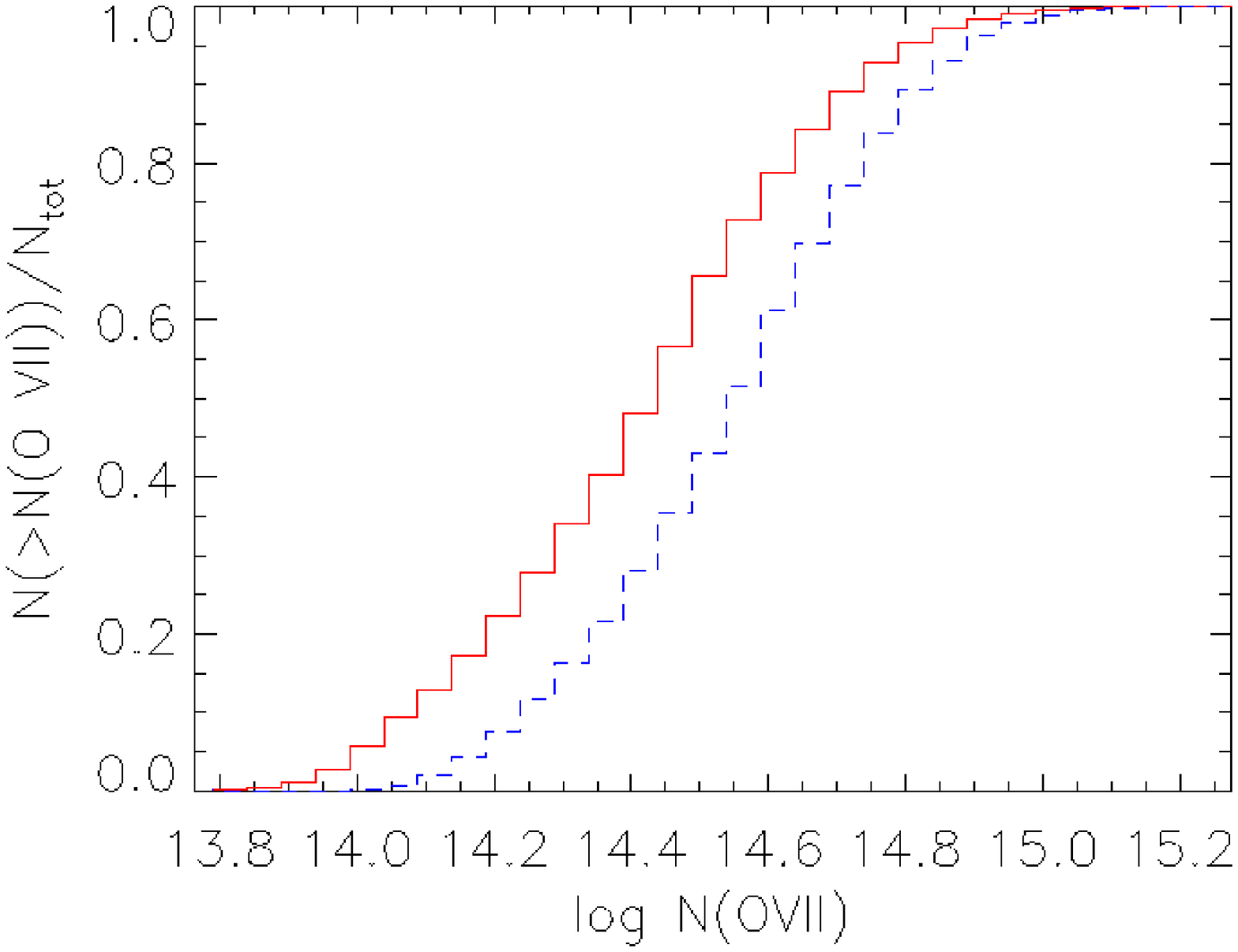,width=\linewidth}
                         \label{hist_2}
                      \end{tabular}
                 \end{center}
             \end{minipage}
        \end{figure*}

Figure \ref{hist} shows the all sky, statistical N(OVII) distributions
for the three galaxies and two types of estimates, employing the weighting
scheme outlined above and figure \ref{hist_2} shows the cumulative histograms of the calculated column densities.

\begin{figure}
           \begin{minipage}{\linewidth}
              \centering
              \caption{Logarithm of N(O VII) in cm$^{-2}$ for the three simulated galaxies, excluding the disc, satellites or HVCs.  From top to bottom: K33, K26, K15}
              \begin{tabular}{c}
                \epsfig{file=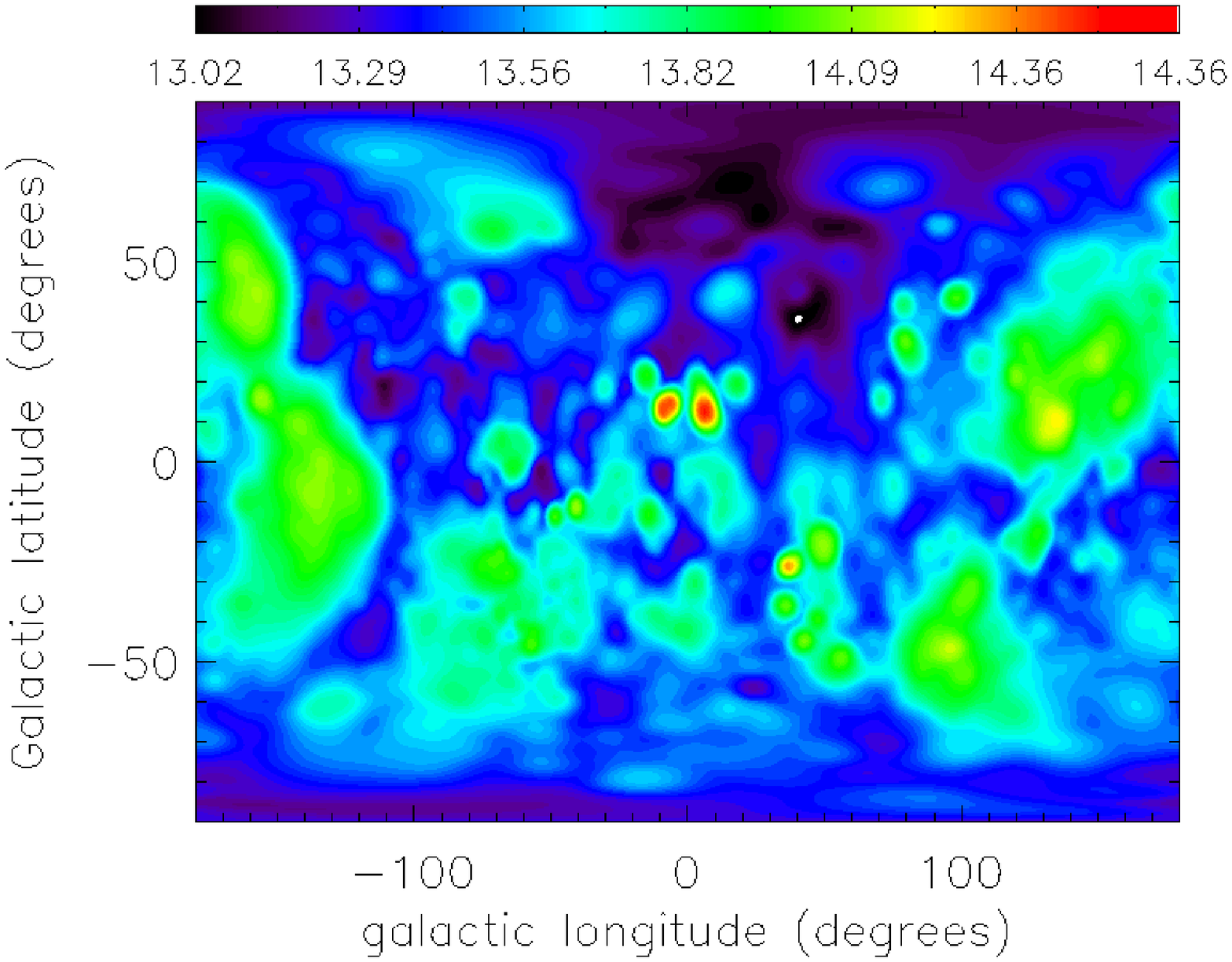,width=\linewidth} \\
                \epsfig{file=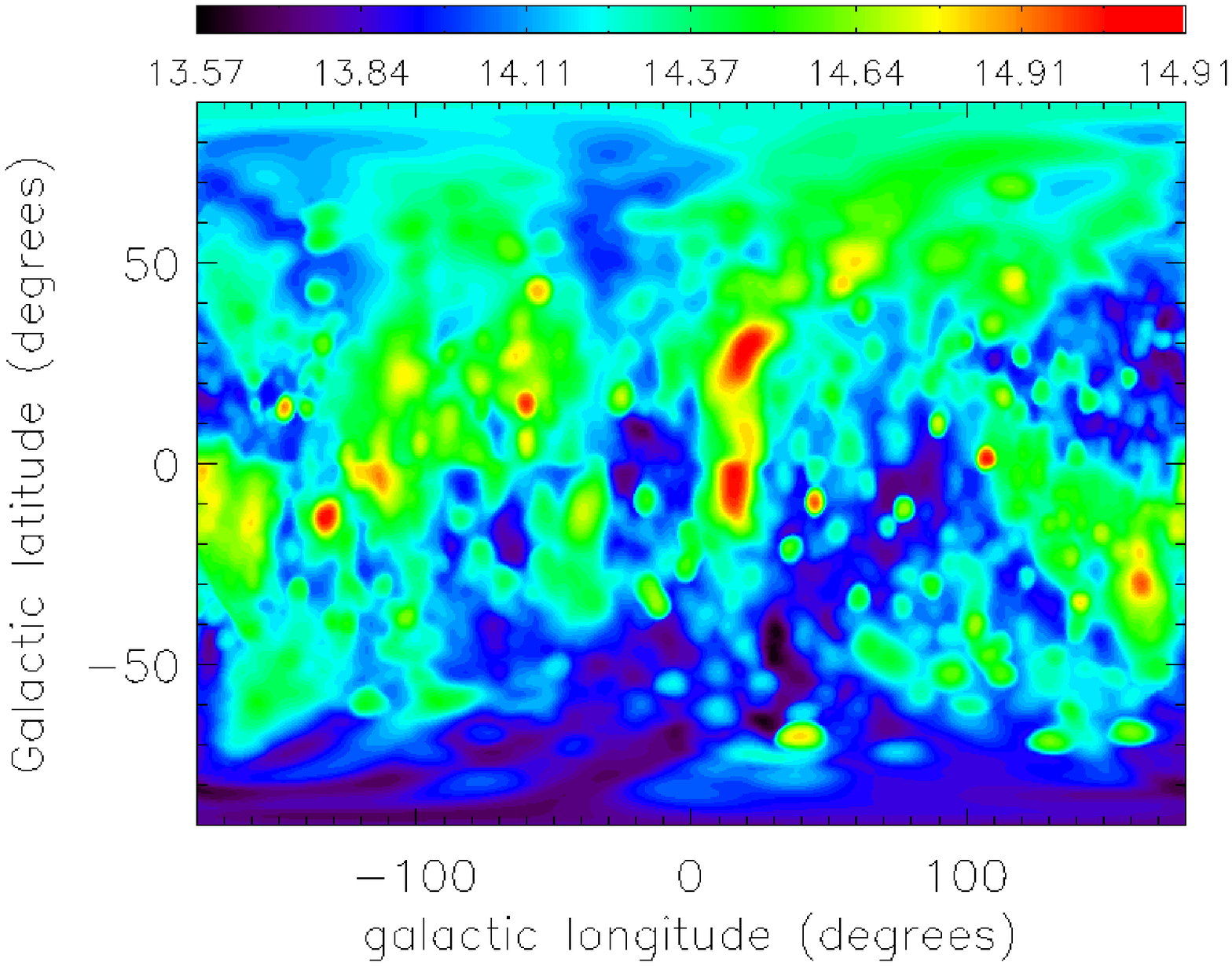,width=\linewidth} \\
                \epsfig{file=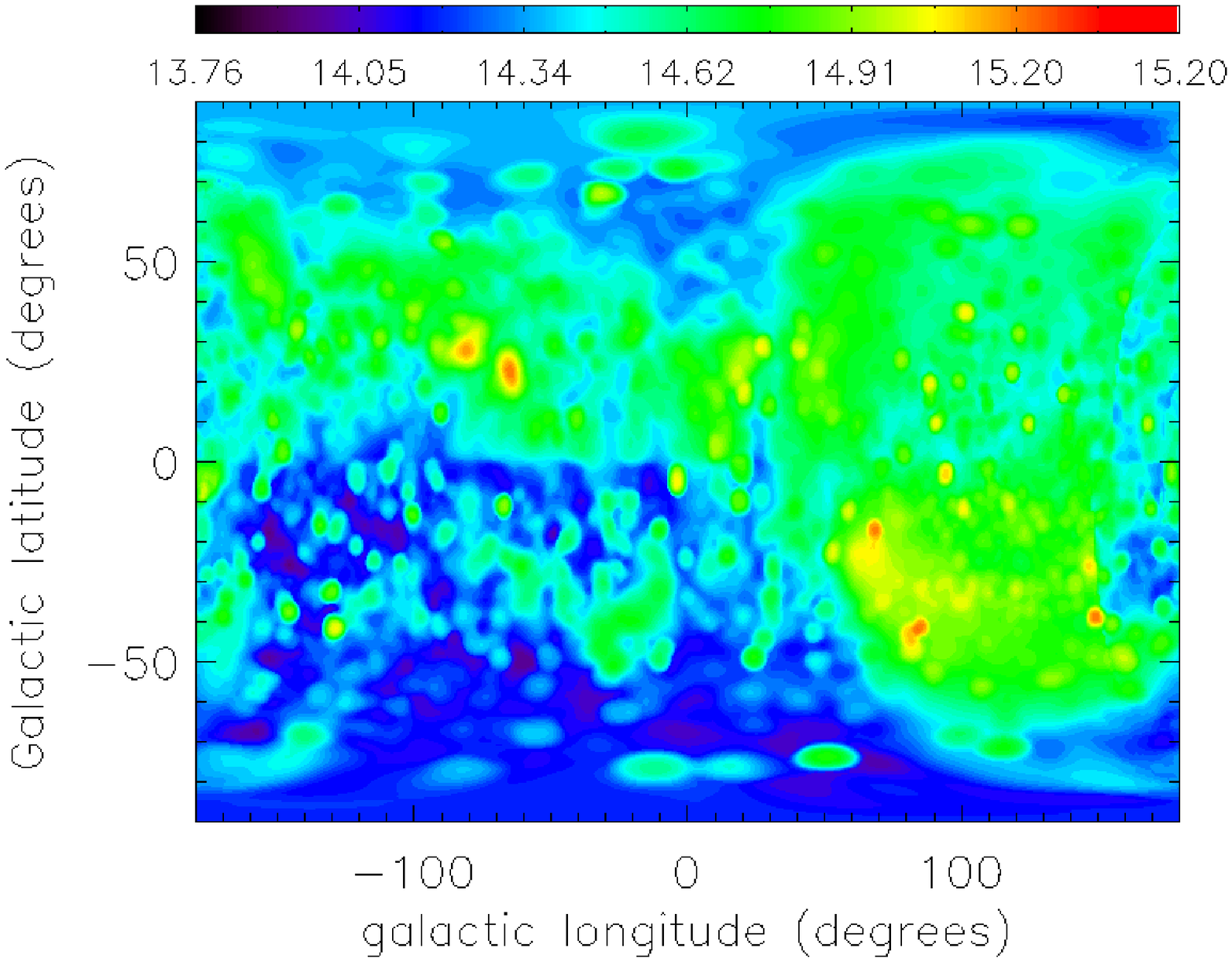,width=\linewidth}
                \label{sky1} 
              \end{tabular}
            \end{minipage}
         \end{figure}

\begin{figure}
           \centering
           \caption{Logarithm of N(O VII) in cm$^{-2}$ for the three simulated galaxies, including 
                    the disc and all objects in the halo.  From top to bottom: K33, K26, K15}
           \begin{tabular}{c}
             \epsfig{file=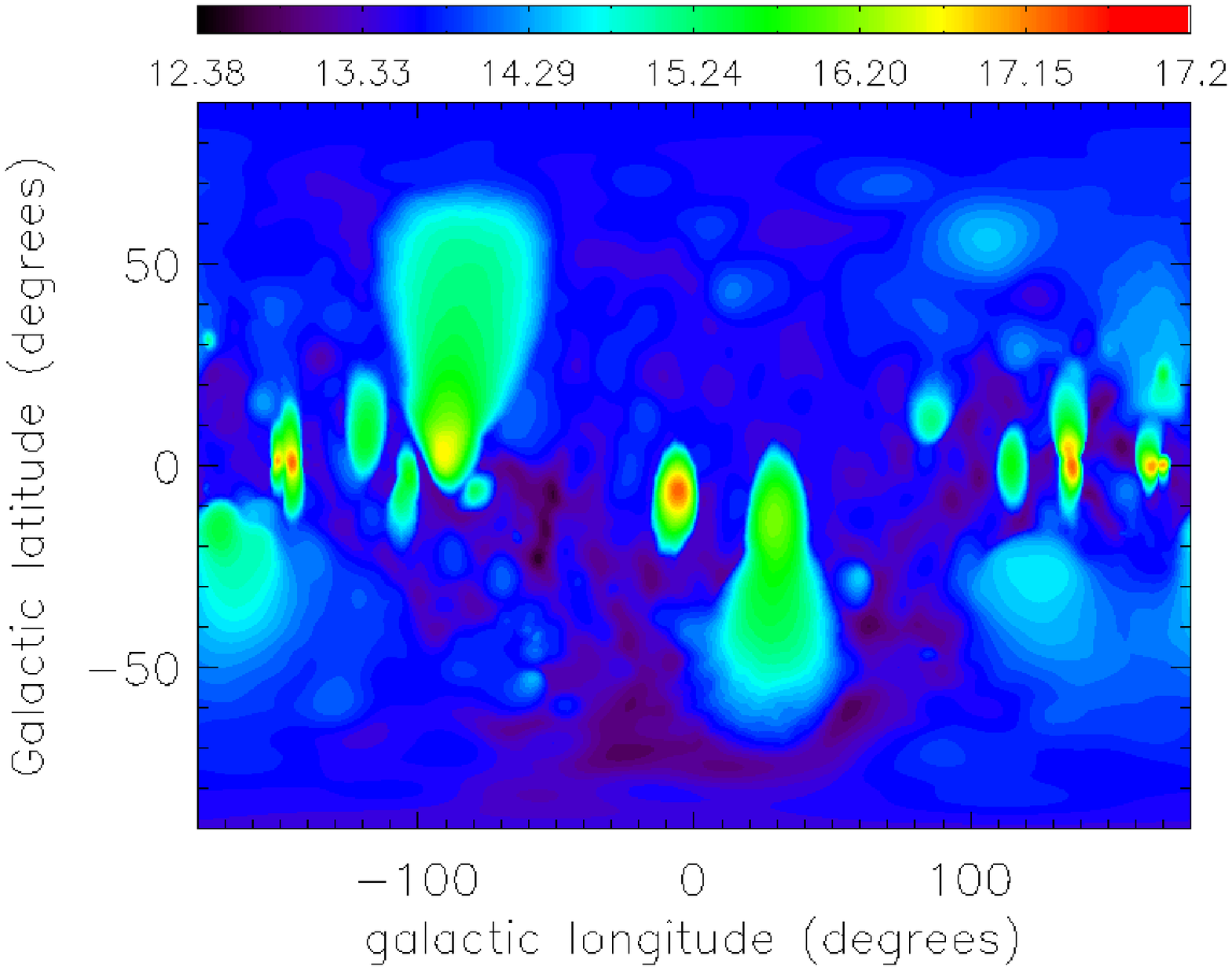,width=\linewidth} \\
             \epsfig{file=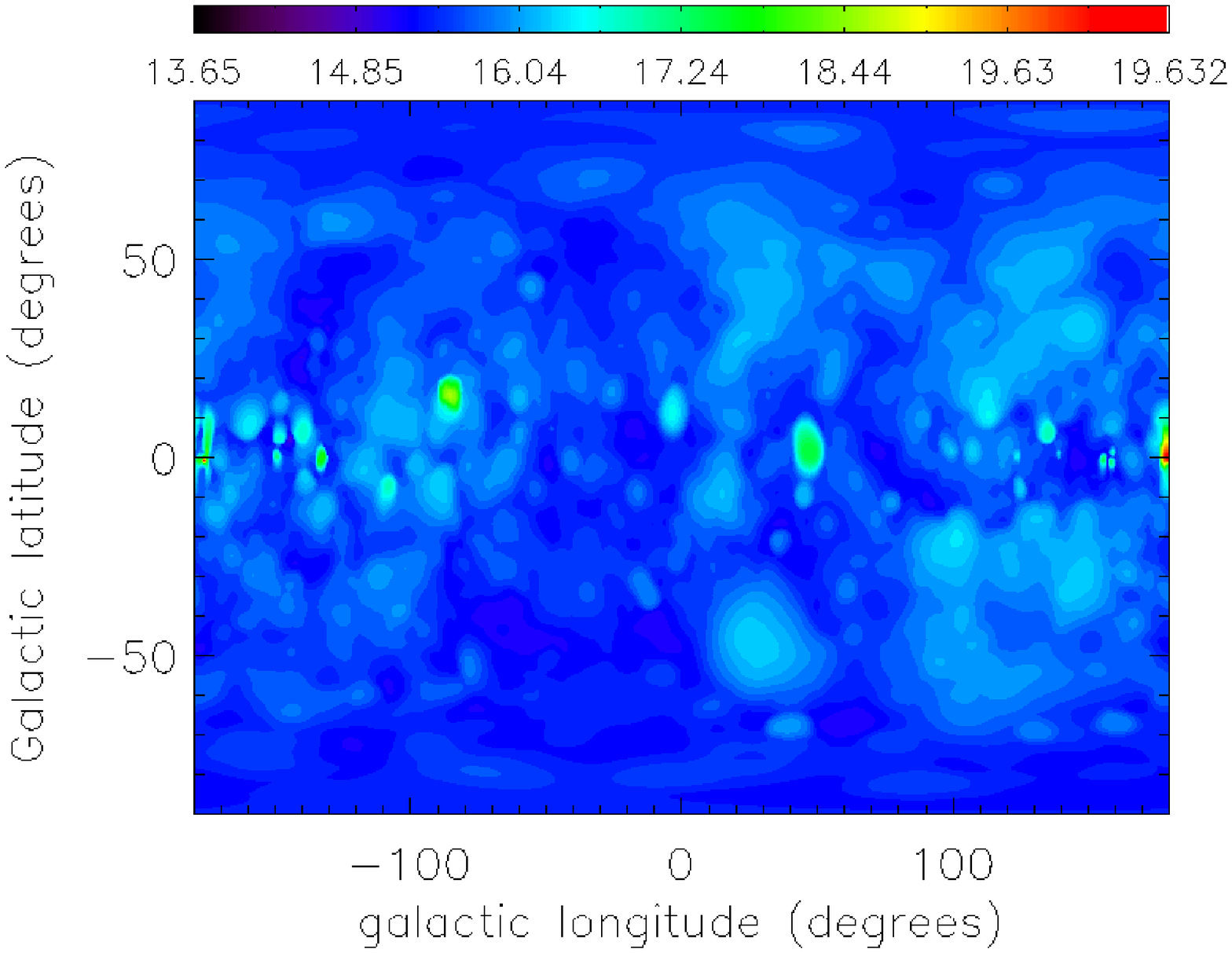,width=\linewidth} \\
             \epsfig{file=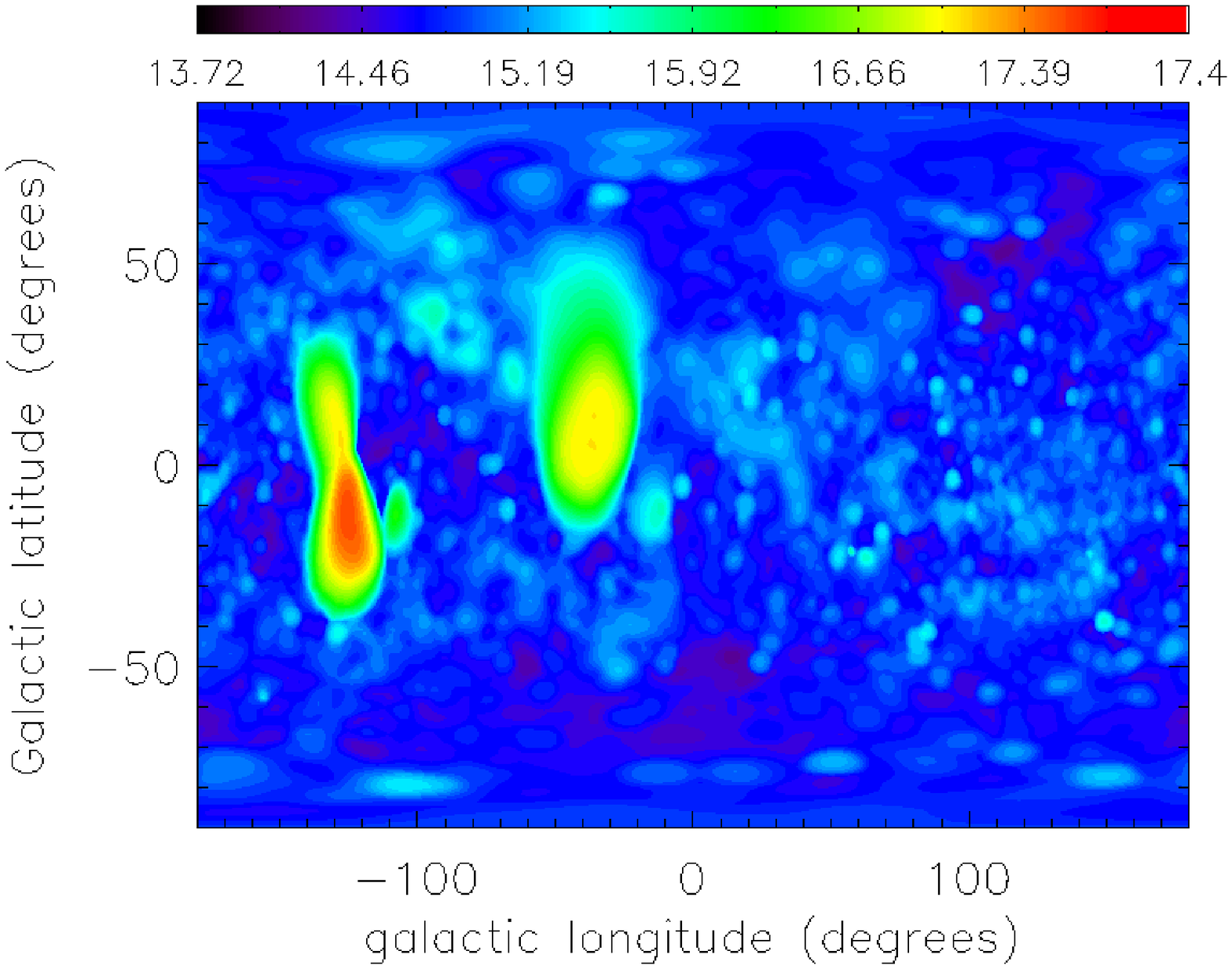,width=\linewidth}
             \label{sky2}
           \end{tabular}
         \end{figure}

Figure \ref{sky1} shows contours of the logarithm of the O~VII column density as a function of galactic longitude and latitude when all disc and ``dense'' 
(see above) halo gas particles are removed from the simulation data. Figure \ref{sky2} shows the results with all the particles included.  Both results have been calculated with a constant one degree distance between the lines of sight.

For all galaxies, the halo only O~VII column density is below the observational upper limit of $\sim 5\cdot10^{15}$ cm$^{-2}$ \citep{yao2008}. 
There is a spread of approximately one order of magnitude in column density values that results from local fluctuations in density and/or metallicity. 

With the inclusion of all SPH particles the mean column density is not significantly affected, but the distributions all display a tail towards higher column densities (see Table \ref{stattable} and figures \ref{hist} and \ref{hist_2}).  The areas near the galactic plane are dominated by disc contribution, while halo objects give extended high column density areas at latitudes away from the galactic plane.  

As can be seen from figure \ref{sky2}, high O~VII column densities are quite common when the disc is included.  This is due to supernova explosions heating the gas and, of course, due to the high metallicity of the disc gas. It is worth noting that a feature near the disc plane of K26 can still be seen in figure \ref{sky1}, when the disc is removed.  This is a typical example of the effects of supernova-driven outflow, driving high-metallicity hot gas into the lower halo.
In the case of K33, the characteristic features in Figure \ref{sky2} come from a large warp surrounding the galaxy, product of the tidal disruption of a satellite, previously accreted onto the galaxy.  

In order to assess how the inclusion of the cosmic UV background affects the results, we repeated the halo-only run using an oxygen fraction table calculated without a UV background.  As illustrated in the left panel of Figure \ref{hist}, the overall effect of the UV field is to shift the distribution to slightly lower column densities.  This happens because the hardest photons of this radiation can ionise oxygen to levels higher than the sixth, so the latter remains relatively underpopulated.  However, this is not the case for K33, where the UV background has the opposite effect, that is to increase the column density values.  This is easily understood, since the average temperature of the K33 halo is somewhat lower, due to its lower mass, so the UV field brings oxygen from lower ionisation stages to the sixth.
        
\begin{figure}
            \centering
            \caption{N(O VII) as a function of the impact parameter b for galaxy K15, which has a characteristic circular speed similar to that of M31.} 

            \epsfig{file=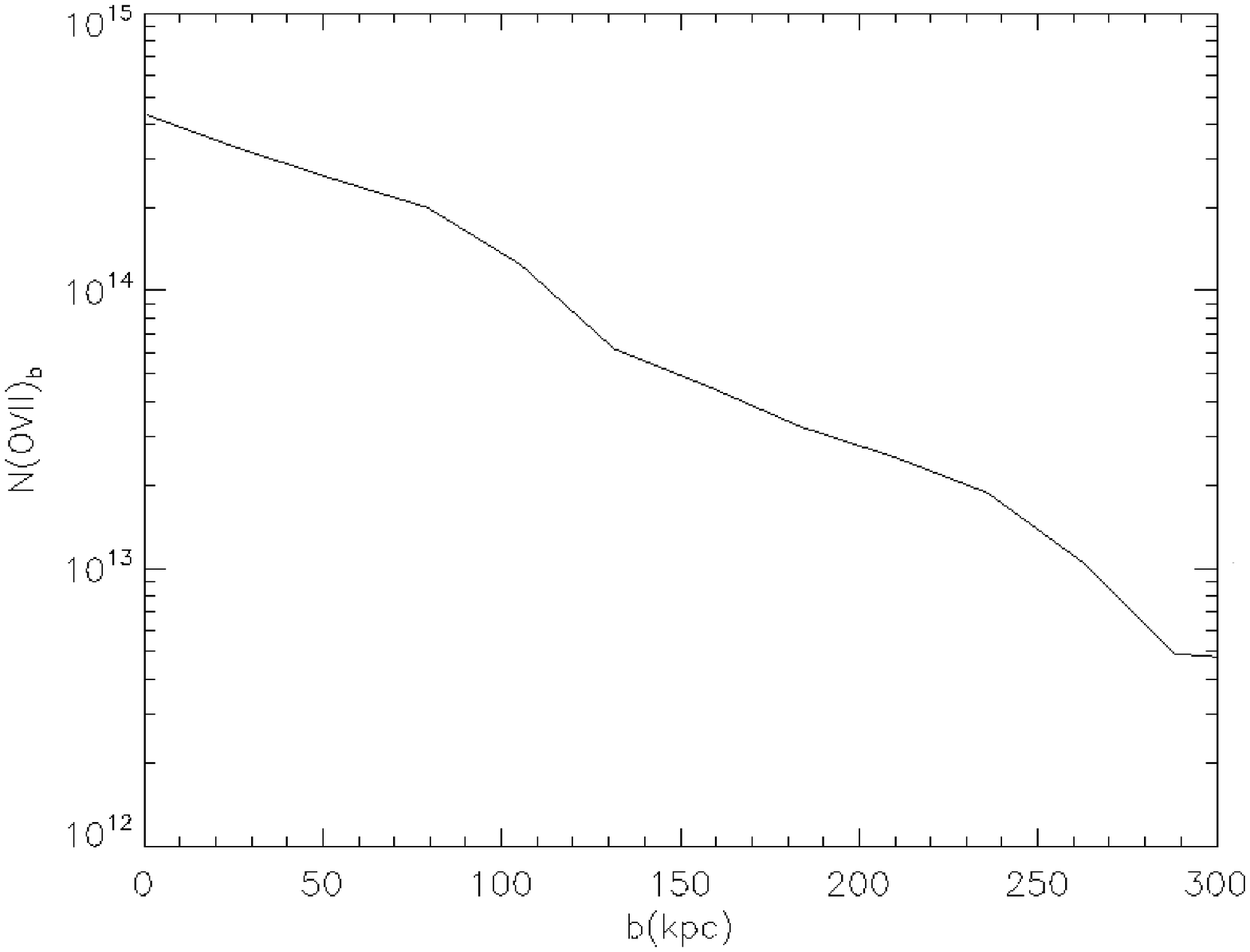, width = \linewidth}
            \label{monte_carlo}
          \end{figure}

The distance of 500kpc at which we decided to stop the integration was chosen so that we would only probe the halo of a single galaxy.  
Integrating to a somewhat smaller distance of 300kpc does not affect the results significantly, owing to the low gas densities at these distances from the galaxy.  For distances larger than 500kpc, one could consider the possibility that a line of sight passes through the haloes of neighbouring galaxies, which would in principle also give a contribution.  To test the effect of a line of sight intersecting the halo of a neighbouring galaxy we calculated the N(OVII) values for 10000 random lines of sight passing through the halo of K15 (this galaxy was chosen because the virial temperature is comparable to that of M31 --- see section 4).  The lines of sight were chosen to originate and end at $r$=500 kpc distance from the central galaxy. Figure \ref{monte_carlo} shows the result of this calculation for impact parameters up to 300kpc.  
For larger impact parameters the contribution drops even further.

It is apparent from Figure \ref{monte_carlo} that, unless a line of sight passes through a neighbouring halo with an impact parameter lower than $\sim$100-200kpc, the contribution from this halo will be negligible.

         \begin{table}
             \begin{minipage}{\linewidth}
                \begin{center}
                \caption{Mean logarithm of column densities $<\log{N(O VII)}>$ for different runs.}
                \begin{tabular*}{\linewidth}{@{}p{0.15\linewidth}p{0.25\linewidth}p{0.25\linewidth}p{0.2\linewidth}@{}}
                   \hline
                    Galaxy     &  Mean $\log{N(O VII)}$ &  
                     Mean $\log{N(O VII)}$ & Mean $\log{N(O VII)}$ \\
                      & without disc  & including disc & without UV or disc \\
                    \hline
                    K33 & 13.55 & 13.72 & 13.46 \\
                    K26 & 14.13 & 14.42 & 14.25 \\
                    K15 & 14.41 & 14.55 & 14.54 \\
                    \hline
                    \label{stattable}
                \end{tabular*}
                \end{center}
             \end{minipage}
          \end{table}
     
        %\clearpage

   \section{Discussion and Conclusions}\label{discussion}
We have in this work determined the O~VII column density of hot halo gas
along various lines-of-sight in three simulated disc galaxies. In order
to emulate observations performed at the solar position in the Milky Way, 
we have, in each galaxy, chosen a position in the midplane of disc, located
8 kpc from the centre of the galaxy. From this position we have drawn
lines-of-sights emulating all sky coverage.

We have tested that the results obtained do not in any significant way
depend on where, along the $R=8$ kpc midplane circle, the observer 
is assumed to be located. Moreover, we have for galaxy K33 tested that
the results obtained do not in any significant way depend on the
numerical resolution of the galaxy formation simulation.

For the two Milky Way sized galaxies, of $V_c=207$ and 245 km/sec, 
respectively, we find median halo gas O~VII column densities of 
1.5-2.5$\cdot10^{14}$ cm$^{-2}$, with a dispersion in log($N$(O~VII)) of
about 0.2. For the somewhat smaller disc galaxy K33, of $V_c=180$ km/s,
we find a median halo gas O~VII column density of 
3$\cdot10^{13}$ cm$^{-2}$, also with a dispersion in log($N$(O~VII)) of
about 0.2. The lower value for this galaxy is primarily due to the
somewhat lower (virial) temperature of the hot gas (by about a
factor of two), which, cf. Fig. \ref{fractions}, leads to a significant 
reduction of the number of oxygen ions in the sixth ionization stage for 
the relevant density range, $n_H \sim 10^{-6}-10^{-3.5}$ cm$^{-3}$.

The main result of the paper is hence that for the two Milky Way sized
galaxies, the median predicted halo O~VII column density lies about a factor of
20 below the observational upper limit of $5\cdot10^{15}$ cm$^{-2}$
reported by \cite{yao2008} (and is comparable to the upper limits of 
$log(N(OVII))<14.2-14.8$ set by \citet{yao2010}, but these limits are
based on halo sight-lines, in general, passing far away from the galactic
centres a comparison is less straightforward --- see also below). Moreover, not a single line-of-sight was
found in any of the galaxies with a halo O~VII column density exceeding
the \cite{yao2008} limit.

When the disc gas, as well as gas in satellites and HI clouds, is not
excluded in the O~VII column density estimates, the median $N$(O~VII)
increases by about a factor of two. Moreover, the $N$(O~VII) distributions 
display
tails towards significantly larger values with a few lines-of-sight
exceeding the \cite{yao2008} limit --- this highlights the importance
of excluding ``contamination'' from the disc and other non-halo
components, as discussed by the same authors. 
It also strongly suggests that the routinely reported ionic column densities of $N(O VII)\sim 10^{16}$cm$^{-2}$
at zero redshift originate in gas located near or inside the galactic disc, while the extended gaseous corona has a much lower contribution.
The largest values of $N$(O~VII)
along a few lines-of-sight are mainly caused by either hot, super-nova
driven bubbles in the disc gas or (in the case of K33) an outer,
partly photo-ionized warp, originating from gas stripped off a 
previously accreted satellite.

To test the sensitivity of our results on the assumed UVB and (simplified)
radiative transfer scheme, we repeated the calculations assuming no
UVB. For the two Milky Way like galaxies, this leads to marginally
larger median $N$(O~VII) values. The reason for this is that at the
(virial) temperatures of the halo gas, O~VII is the most abundant
oxygen state, and with the UVB switched on, some of the O~VII ions
are photo-ionized to (mainly) the O~VIII stage, lowering the population
of O~VII ions. For the lower mass galaxy K33, on the other hand,
the effect of the UVB is to boost the O~VII ion population.  
This is confirmed by calculations (not presented in this paper)
of the O~VI and O~VIII column densities for the same lines of sight.

Finally, we have tried to quantify the effect a neighbouring galactic halo might have 
on the O~VII column density for a random line of sight.  
The probability of that happening and the relative effect this might have on the final result
 depends, of course, on the sizes of the two galaxies and on the distance between them.
In a simple calculation, we have randomly chosen the starting and ending coordinates 
of 10000 lines of sight outside a certain halo and calculated their O~VII column densities. We find that the contribution is very small
unless the impact parameter of the line of sight to
the centre of the galaxy is lower than 100-200kpc.  
Including the disc gas and satellites in this case does not alter the results presented above
 unless the impact parameter is very small (smaller than 50kpc).
This is in agreement with what \citet{bregmanlloyd07} find for their line of sight passing closest to M31, at a distance of 380kpc, giving no substantial contribution to the absorption. 

Considering this result, possible contamination to the lines of sight that probe the Milky Way halo could likely only come from M31
(more distant galaxies would have their O VII line redshifted, so it would be relatively easy to distinguish their halo gas contribution from the one coming from local gas; this is beyond the scope of this paper, and will be the topic of a forthcoming paper --- see also below).
M31 is located at a distance of approximately 700kpc from our galaxy, which would cause it to 
contribute to the O VII column density with more than $10^{14}$cm$^{-2}$ for only about $0.5\%$ of the sky area ($b\la 100$kpc --- see Fig.~\ref{monte_carlo}). Excluding $b\le 200$kpc would
reduce the M31 contribution to less than about $2\times10^{13}$cm$^{-2}$, and
still only exclude about 2\% of the sky area.

As mentioned above, it will be the topic of a forthcoming paper to investigate the contribution from the haloes of other, more distant galaxies. This study
will be based on the proper statistical descriptors at the large-scale
galaxy distribution. 

In conclusion, the present observational upper limit is perfectly
consistent with the results of state-of-the-art galaxy formation 
models, based on fully cosmological simulations. Moreover,
although the predicted halo $N$(O~VII) values lie below the
observational upper limit, the difference is still only about an
order of magnitude. With more sensitive future 
X-ray observatories, such as IXO, which will provide an order of magnitude 
increase in collecting area and a very high spectral resolution, 
it is very likely that one will be able detect O~VII in the halo of
the Milky Way, and possibly also in the haloes of other galaxies.

   \section*{Acknowledgments}
We gratefully acknowledge abundant access to the computing facilities
provided by the Danish Centre for Scientific Computing (DCSC). This
work was supported by the DFG Cluster of Excellence ``Origin and Structure
of the Universe''. The Dark Cosmology Centre is funded by the Danish
National Research Foundation.

\bibliographystyle{mn2e}
\bibliography{ovii}

\begin{thebibliography}{}

\bibitem[\protect\citeauthoryear{{Asplund}, {Grevesse} \& {Sauval}}{{Asplund}
  et~al.}{2005}]{asplund05}
{Asplund} M.,  {Grevesse} N.,    {Sauval} A.~J.,  2005, in {Barnes} III T.~G.,
  {Bash} F.~N.,  eds, Cosmic Abundances as Records of Stellar Evolution and
  Nucleosynthesis Vol.~336 of Astronomical Society of the Pacific Conference
  Series, {The Solar Chemical Composition}.
pp 25--+

\bibitem[\protect\citeauthoryear{{Bregman} \& {Lloyd-Davies}}{{Bregman} \&
  {Lloyd-Davies}}{2007}]{bregmanlloyd07}
{Bregman} J.~N.,  {Lloyd-Davies} E.~J.,  2007, ApJ, 669, 990

\bibitem[\protect\citeauthoryear{Br\"{u}ns, Kerp, Kalberla \& Mebold}{Br\"{u}ns
  et~al.}{2000}]{bruns2000}
Br\"{u}ns C.,  Kerp J.,  Kalberla P.,    Mebold U.,  2000, A\&AS, 357, 120

\bibitem[\protect\citeauthoryear{{Fang}, {Mckee}, {Canizares} \&
  {Wolfire}}{{Fang} et~al.}{2006}]{fang06}
{Fang} T.,  {Mckee} C.~F.,  {Canizares} C.~R.,    {Wolfire} M.,  2006, ApJ,
  644, 174

\bibitem[\protect\citeauthoryear{Ferland, Korista, Verner, Ferguson, Kingdon \&
  Verner}{Ferland et~al.}{1998}]{ferland98}
Ferland G.~J.,  Korista K.,  Verner D.,  Ferguson J.,  Kingdon J.,    Verner
  E.,  1998, PASP, 110, 761

\bibitem[\protect\citeauthoryear{{Ganguly}, {Sembach}, {Tripp} \&
  {Savage}}{{Ganguly} et~al.}{2005}]{ganguly05}
{Ganguly} R.,  {Sembach} K.~R.,  {Tripp} T.~M.,    {Savage} B.~D.,  2005, ApJS,
  157, 251

\bibitem[\protect\citeauthoryear{Grcevich \& Putman}{Grcevich \&
  Putman}{2009}]{GP09}
Grcevich J.,  Putman M.,  2009, ApJ, 696, 385

\bibitem[\protect\citeauthoryear{Haardt \& Madau}{Haardt \& Madau}{1996}]{HM96}
Haardt F.,  Madau P.,  1996, ApJ, 461, 20

\bibitem[\protect\citeauthoryear{{Haardt} \& {Madau}}{{Haardt} \&
  {Madau}}{1997}]{hm1997}
{Haardt} F.,  {Madau} P.,  1997, in {Wickramasinghe} D.~T.,  {Bicknell} G.~V.,
   {Ferrario} L.,  eds, IAU Colloq. 163: Accretion Phenomena and Related
  Outflows Vol.~121 of Astronomical Society of the Pacific Conference Series,
  {The Intrinsic UV/Soft X-Ray Spectrum Of Quasars}.
pp 711--+

\bibitem[\protect\citeauthoryear{Mastropietro}{Mastropietro}{2009}]{M09}
Mastropietro C.,  2009, in The Magellanic System: Stars, Gas, and Galaxies
  Vol.~256 of Proceedings of the International Astronomical Union, IAU
  Symposium,, {Modeling a high velocity LMC: The formation of the Magellanic
  Stream}.
pp 117--+

\bibitem[\protect\citeauthoryear{Nicastro, Mathur, Elvis, Drake, Fang,
  Fruscione, Krongold, Marshall, Williams \& Zezas}{Nicastro
  et~al.}{2005}]{nicastro05}
Nicastro F.,  Mathur S.,  Elvis M.,  Drake J.,  Fang T.,  Fruscione A.,
  Krongold Y.,  Marshall H.,  Williams R.,    Zezas A.,  2005, Nat, 433, L495

\bibitem[\protect\citeauthoryear{{Nicastro}, {Zezas}, {Drake}, {Elvis},
  {Fiore}, {Fruscione}, {Marengo}, {Mathur} \& {Bianchi}}{{Nicastro}
  et~al.}{2002}]{nicastro02}
{Nicastro} F.,  {Zezas} A.,  {Drake} J.,  {Elvis} M.,  {Fiore} F.,  {Fruscione}
  A.,  {Marengo} M.,  {Mathur} S.,    {Bianchi} S.,  2002, ApJ, 573, 157

\bibitem[\protect\citeauthoryear{Peek, Putman \& Sommer-Larsen}{Peek
  et~al.}{2008}]{peek2008}
Peek J.,  Putman M.,    Sommer-Larsen J.,  2008, ApJ, 674, 227

\bibitem[\protect\citeauthoryear{{Rasmussen}, {Kahn}, {Paerels}, {den Herder}
  \& {de Vries}}{{Rasmussen} et~al.}{2003}]{rasmussen03}
{Rasmussen} A.,  {Kahn} S.~M.,  {Paerels} F.,  {den Herder} J.,    {de Vries}
  C.,  2003, in Bulletin of the American Astronomical Society Vol.~35 of
  Bulletin of the American Astronomical Society, {Observational limits to
  highly ionized absorption systems in the intergalactic medium for z<0.15.}.
pp 605--+

\bibitem[\protect\citeauthoryear{{Rasmussen}, {Sommer-Larsen}, {Pedersen},
  {Toft}, {Benson}, {Bower} \& {Grove}}{{Rasmussen} et~al.}{2009}]{rasmussen09}
{Rasmussen} J.,  {Sommer-Larsen} J.,  {Pedersen} K.,  {Toft} S.,  {Benson} A.,
  {Bower} R.~G.,    {Grove} L.~F.,  2009, ArXiv e-prints

\bibitem[\protect\citeauthoryear{{Saha}}{{Saha}}{1921}]{saha1921}
{Saha} M.~N.,  1921, Royal Society of London Proceedings Series A, 99, 135

\bibitem[\protect\citeauthoryear{{Sembach}, {Wakker}, {Savage}, {Richter},
  {Meade}, {Shull}, {Jenkins}, {Sonneborn} \& {Moos}}{{Sembach}
  et~al.}{2003}]{sembach03}
{Sembach} K.~R.,  {Wakker} B.~P.,  {Savage} B.~D.,  {Richter} P.,  {Meade} M.,
  {Shull} J.~M.,  {Jenkins} E.~B.,  {Sonneborn} G.,    {Moos} H.~W.,  2003,
  ApJS, 146, 165

\bibitem[\protect\citeauthoryear{Shull, Jones, Danforth \& Collins}{Shull
  et~al.}{2009}]{Sh.09}
Shull J.,  Jones J.,  Danforth C.,    Collins J.,  2009, ApJ, 700, in press

\bibitem[\protect\citeauthoryear{Sommer-Larsen}{Sommer-Larsen}{2006}]{SL06}
Sommer-Larsen J.,  2006, ApJL, 644, 1

\bibitem[\protect\citeauthoryear{Sommer-Larsen, G\"otz \&
  Portinari}{Sommer-Larsen et~al.}{2003}]{SLGP03}
Sommer-Larsen J.,  G\"otz M.,    Portinari L.,  2003, ApJ, 596, 47

\bibitem[\protect\citeauthoryear{Springel \& Hernquist}{Springel \&
  Hernquist}{2002}]{SH02}
Springel V.,  Hernquist L.,  2002, MNRAS, 649, 649

\bibitem[\protect\citeauthoryear{{Stanimirovi{\'c}}, {Hoffman}, {Heiles},
  {Douglas}, {Putman} \& {Peek}}{{Stanimirovi{\'c}} et~al.}{2008}]{stanim2008}
{Stanimirovi{\'c}} S.,  {Hoffman} S.,  {Heiles} C.,  {Douglas} K.~A.,  {Putman}
  M.,    {Peek} J.~E.~G.,  2008, ApJ, 680, 276

\bibitem[\protect\citeauthoryear{Toft, Rasmussen, Sommer-Larsen \&
  Pedersen}{Toft et~al.}{2002}]{toft2002}
Toft S.,  Rasmussen J.,  Sommer-Larsen J.,    Pedersen K.,  2002, MNRAS, 335,
  799

\bibitem[\protect\citeauthoryear{{Wakker}, {Savage}, {Sembach}, {Richter} \&
  {Fox}}{{Wakker} et~al.}{2005}]{wakker05}
{Wakker} B.~P.,  {Savage} B.~D.,  {Sembach} K.~R.,  {Richter} P.,    {Fox}
  A.~J.,  2005, in {Braun} R.,  ed., Extra-Planar Gas Vol.~331 of Astronomical
  Society of the Pacific Conference Series, {High-velocity O VI in and near the
  Milky Way}.
pp 11--+

\bibitem[\protect\citeauthoryear{Yao, Nowak, Wang, Schulz \& Canizares}{Yao
  et~al.}{2008}]{yao2008}
Yao Y.,  Nowak M.,  Wang Q.,  Schulz N.,    Canizares C.,  2008, ApJ, 672, L21

\bibitem[\protect\citeauthoryear{{Yao}, {Wang}, {Penton}, {Tripp}, {Shull} \&
  {Stocke}}{{Yao} et~al.}{2010}]{yao2010}
{Yao} Y.,  {Wang} Q.~D.,  {Penton} S.~V.,  {Tripp} T.~M.,  {Shull} J.~M.,
  {Stocke} J.~T.,  2010, ApJ, 716, 1514

\end{thebibliography}

\label{lastpage}

\end{document}